\documentclass[twoside]{article}
\usepackage{amsmath,amsfonts,amssymb,latexsym}
\usepackage[english]{babel}
\usepackage{epsfig,graphicx}
\usepackage{indentfirst,cite}
\newcommand{\mytitle}[1]{\large \sc #1 \\}
\newcommand{\avtor}[1]  {\large \it #1 \\}

mm  \leftmargin 5pt
\newtheorem{assum}{Assumption}
\newtheorem{lem}{Lemma}
\newtheorem{demo}{Statement}
\newtheorem{rem}{Remark}
\newtheorem{teo}{Theorem}
\newtheorem{prop}{}
\newtheorem{defin}{Definition}

\newtheorem{sled}{Corollary}

\makeatletter \@addtoreset{equation}{section}
\@addtoreset{prop}{section}
\@addtoreset{sled}{teo} \makeatother

\makeatother 

\begin{document}
\begin{center}

\mytitle{ Quasi-energy spectral series and the Aharonov-Anandan
phase for the nonlocal Gross--Pitaevsky equation }

\bigskip
\avtor
{Lisok $^*$\footnote{e-mail: lisok@mph.phtd.tpu.edu.ru} A.L.,
Trifonov$^*$\footnote{e-mail:  trifonov@phtd.tpu.edu.ru} A.Yu.,
Shapovalov$^{\#}${\footnote {e-mail: shpv@phys.tsu.ru} A.V.}}

$^{*}$ {\it Laboratory of Mathematical Physics\\
Mathematical Physics Department,\\
Tomsk Polytechnical University,\\
Lenin ave., 30, 634050, Russia }\vskip 0.4cm
$^{\#}$ {\it Theoretical Physics Department,\\
Tomsk State University, \\
Lenin ave., 36, 634050, Russia}
\end{center}

\begin{abstract}
For the nonlocal $T$-periodic Gross--Pitaevsky operator, formal
solutions of the Floquet problem asymptotic in small parameter
$\hbar$, $\hbar\to0$, up to $O(\hbar^{3/2})$ have been
constructed. The quasi-energy spectral series found correspond to
the closed phase trajectories of the Hamilton--Ehrenfest system
which are stable in the linear approximation. The monodromy
operator of this equation has been constructed to within $\hat
O(\hbar^{3/2})$ in the class of trajectory-concentrated functions.
The Aharonov--Anandan phases have been calculated for the
quasi-energy states.
\end{abstract}

\section{Introduction}

The global properties of the solutions of mathematical physics
equations which simulate a physical system are determined by the
nontrivial geometry and topology of the system. A basic problem in
investigating the behavior of a system is to find mathematical
structures which would  describe the topology of the system and make
possible its efficient analysis. For quantum systems subjected to an
external action cyclically varying with time, this structures are
well known as topological or geometric phases (GP's) of the wave
function. Some manifestations of GP's were also observed in
polarization optics and mechanics. Theoretical and experimental
studies of geometric phases in quantum mechanics have been performed
since the late 20s of the last century. The work by Berry
\cite{shapovalov:BERRY} has substantially extended the area of
application of GP's and has led to the interpretation of this notion
in terms of gauge symmetry and an effective gauge field or, in
geometric formulation, in terms of a Hilbert bundle with a
finite-dimensional base. A detailed description of this problem can
be found, for instance, in \cite{shapovalov:VINI,
shapovalov:KLYSHKO,Moore}.

The theory of GP's in quantum mechanics is based on the linearity
of the Schr\"odinger equation. For nonlinear equations, the notion
of GP can also be introduced (see, e.g., \cite{shapovalov:VINI}),
but the GP's in nonlinear systems are less well understood.
Difficulties arise not only for lack of the principle of
superposition of solutions. In case of a nonlinear equation, the
expression for the GF's is constructed by analogy with the linear
theory (see, e.g., \cite{shapovalov:VINI}), and it is not fully
obvious that the expression constructed is governed only by the
geometry of the system and that it does not involve the dynamic
contribution due to nonlinearity. The nontrivial topology of a
system can be determined by the boundary conditions as well as by
the external fields. The latter appear in the equation as variavle
factors. Few, if any methods are available to construct exact
solutions for equations of this type. Therefore, it is natural to
study GP's in nonlinear systems by using adequate approximation
methods.

In this connection, the analysis of the GP's for the nonlocal
Gross--Pitaevsky equation given below would be of interest. The
evolution of the initial state of the nonlinear equation
(\ref{tr1.1}) in the semiclassical approximation is determined in
fact by a set of associated linear Schr\"odinger equations. This
allows one not only to calculate in full measure the GP for the
nonlinear equation in explicit form, but also to interprete it in
terms of ``an induced gauge field'' (see \cite{shapovalov:VINI}),
which in turn proves its geometric origin.

A class of solutions asymptotic in small parameter $\hbar\to 0$,
which are localized in a neighborhood of some phase curve, has
been constructed
\cite{shapovalov:BTS1,shapovalov:BTS2,lst3,lst7,lst_sigma} for the
nonlocal Gross--Pitaevsky equation (NGPE) with an external field
(variable coefficients) and a nonlocal nonlineariry
\begin{equation}
\label{tr1.1} \{ -i\hbar\partial_t +\widehat {\mathcal
H}(t)+\varkappa\widehat V(t,\Psi)\}\Psi=0, \qquad \Psi\in
L_2({\mathbb R}^n_x).
\end{equation}

Here, the operators
\begin{gather}
\widehat{\mathcal H}(t)={\mathcal H}(\hat z,t), \label{phaz_lst1.2a}\\
\widehat V(t,\Psi )=\int\limits_{{\mathbb R}^n} d\vec
y\,\Psi^*(\vec y ,t) V(\hat z,\hat w,t)\Psi (\vec y ,t)
\label{phaz_lst1.3a}
\end{gather}
are functions of the noncommuting and Weyl-ordered operators
\[
\hat z=\Big( -i\hbar\frac\partial{\partial\vec x}, \vec x\Big),
\qquad \hat w=\Big( -i\hbar\frac\partial{\partial\vec y}, \vec
y\Big), \qquad \vec x,\vec y\in {\mathbb R}^n,
\]
the function $\Psi^*$ is complex conjugate to $\Psi$, $\varkappa$
is a real parameter, and $\hbar$ is a ``small parameter'',
$\hbar\in[0,1[$. For the operators $\hat z$ and $\hat w$ the
following commutation relations are valid:
\begin{eqnarray}
&[\hat z_k,\hat z_j]_-=[\hat w_k,\hat w_j]_-=i\hbar J_{kj},\quad 
[\hat z_k,\hat w_j]_- = 0,\qquad
k,j=\overline{1,2n},
\end{eqnarray}
where $J =\|J_{kj}\|_{2n\times 2n}$ is a simplectic unit matrix
$$ J=\begin{pmatrix}0&-{\mathbb I}\\
{\mathbb I}&0\end{pmatrix}_{2n\times 2n}.
$$

The goal of this work is to construct explicit expressions for the
quasi-energy spectral series corresponding to the closed phase
trajectories of the Hamilton--Ehrenfest system which are stable in
the linear approximation and GP's in the class of asymptotic
solutions of equation (\ref{tr1.1})-(\ref{phaz_lst1.3a}),
constructed in
\cite{shapovalov:BTS1,shapovalov:BTS2,lst3,lst7,lst_sigma}, for the
case where the operators $\widehat{\mathcal H}(t)$ and $\widehat
V(t,\Psi)$ are $T$-periodic functions of time:
\begin{equation}
\widehat{\mathcal H}(t+T)=\widehat{\mathcal H}(t),\qquad
  \widehat V(t+T,\Psi)=\widehat V(t,\Psi).\label{1.period}
\end{equation}



For the linear Schr\"odinger equation($\varkappa =0$) with a
$T$-periodic Hamiltonian $\widehat{\mathcal H}(t)$, Zeldovich
\cite{shapovalov:ZELDOVICH} and Ritus \cite{shapovalov:RITUS} were
first to introduce an important class of solutions
--- quasi-energy states $\Psi_{\mathcal E}(\vec
x,t,\hbar)$, which possess the property
\begin{equation}
\Psi_{\mathcal E}(\vec x,t+T,\hbar)=e^{-i{\mathcal E} T/\hbar}
\Psi_{\mathcal E}(\vec x,t,\hbar).\label{tr1.2aa}
\end{equation}
The quantity ${\mathcal E}$, involved in (\ref{tr1.2aa}), is
called quasi-energy, and it is defined modulo $\hbar\omega$
($\omega = 2\pi/T$), i.e. ${\mathcal E}'={\mathcal
E}+m\hbar\omega$, $m\in{\mathbb Z}$. States of this type play the
key part in describing quantum-mechanical systems under strong
periodic external actions for which the conventional methods of
nonstationary perturbation theory are not applicable.

For the nonlocal Gross--Pitaevsky equation
(\ref{tr1.1})-(\ref{phaz_lst1.3a}), we shall set the problem of
finding a quasi-energy spectrum in the form (\ref{tr1.2aa}).


The quasi-energy states (\ref{tr1.2aa}) are a particular case of
the cyclic states introduced by Aharonov and Anandan
\cite{shapovalov:ANANDAN} (see also \cite{shapovalov:VINI}). By
cyclic evolution of a quantum system on a time interval [$0,T$] is
implied the following: the state vector $\Psi(t)$ has the form
\begin{equation}
\Psi(t)=e^{if(t)}\varphi (t), \qquad t \in [0,T], \label{tr1.4}
\end{equation}
where
\begin{eqnarray}
& f(T)-f(0) = \phi (\bmod~2\pi), \label{tr1.5} 
\quad\varphi(T)=\varphi(0). 
\end{eqnarray}
The total phase $\phi$ of the function(\ref{tr1.4}) is subdivided
into two terms: the dynamic phase
\begin{equation}
\delta =-\frac1\hbar \int\limits_0^T dt \frac{\langle\Psi(t)|
[\widehat{\mathcal H}(t)+\varkappa\widehat
V(t,\Psi(t))]|\Psi(t)\rangle}{\langle\Psi(t)|\Psi(t)\rangle}\label{tr1.7}
\end{equation}
and the Aharonov--Anandan geometric phase
\begin{equation}
\gamma =i\int\limits_0^T dt \frac{\langle \varphi(t)|
\dot\varphi(t)\rangle} {\langle\varphi(t)|\varphi(t)\rangle}.
\label{tr1.8}
\end{equation}
Comparing (\ref{tr1.2aa}) and (\ref{tr1.4}), we obtain that for
the quasi-energy states the function $f(t)$ is given by
\begin{equation}
f(t) = -{\mathcal E} t/\hbar, \label{tr1.9}
\end{equation}
while for the total phase $\phi$, according to (\ref{tr1.5}), we
have
\begin{equation}
\phi=-\frac{{\mathcal E} T}\hbar \quad
(\bmod\,2\pi).\label{tr1.10}
\end{equation}
By virtue of (\ref{tr1.7})--(\ref{tr1.10}), the Aharonov--Anandan
phase $\gamma_{\mathcal E}$ corresponding to a given quasi-energy
state $\Psi_{\mathcal E}(\vec x,t,\hbar)$ can be determined by
formula $(\bmod\,2\pi)$:
\begin{equation}
\gamma_{\mathcal E} = -\frac{{\mathcal E} T}\hbar +\frac1\hbar
\int\limits_0^T dt \frac{\langle \Psi_{\mathcal E}|
[\widehat{\mathcal H}(t)+ \varkappa\widehat
V(t,\Psi(t))]|\Psi_{\mathcal E}\rangle} {\langle \Psi_{\mathcal E}
|\Psi_{\mathcal E} \rangle}.\label{tr1.11}
\end{equation}

\section[Statement of the problem in the class of semiclassically concentrated functions]
{Statement of the problem in the class of semiclassically \\
concentrated functions}

We shall construct asymptotic solutions for the nonlocal
Gross--Pitaevsky equation with the following assumption for the
Weyl symbols of the operators ${\mathcal H}(\hat z,t)$ and $V(\hat
z,\hat w,t)$ in (\ref{tr1.1})-(\ref{phaz_lst1.3a}):

\begin{assum} \label{assum1}
The functions ${\mathcal H}(z,t)$ and $V(z,w,t)$ are
$C^\infty$-smooth functions and for any multiindices $\alpha$ and
$\beta$, $(\alpha,\beta\in{\mathbb Z}_+^{2n})$ and $T>0$, there
exist constants $\kappa>0$, $C_\alpha(T)$, and
$C_{\alpha\beta}(T)$, such that
\[ \Big|\frac{\partial^{|\alpha|}{\mathcal H}(z,t)}{\partial z^\alpha}\Big| \le
C_\alpha(T)(1+|z|)^\kappa, \quad \Big|
\frac{\partial^{|\alpha+\beta|}V(z,w,t)} {\partial z^\alpha
\partial w^\beta}\Big|\le C_{\alpha\beta}(T)(1+|z|+|w|)^\kappa,
 \quad z,w\in {\mathbb R}^{2n}, \quad 0\le t\le T. \]
\end{assum}
Here,
\begin{gather*}
\frac{\partial^{|\alpha|}V(z)}{\partial z^\alpha}=
 \frac{\partial^{|\alpha|}V(z)}{\partial z_1^{\alpha_1}
 \partial z_2^{\alpha_2}\dots \partial z_{2n}^{\alpha_{2n}}},
 \quad \alpha_j\in{\mathbb Z}_+, \quad j=\overline{1,2n},\\
\alpha=(\alpha_1,\alpha_2,\dots,\alpha_{2n}), \quad
 |\alpha|=\alpha_1+\alpha_2+\dots+\alpha_{2n}, \quad
 z^\alpha=z_1^{\alpha_1} z_2^{\alpha_2}\dots  z_{2n}^{\alpha_{2n}}.
\end{gather*}

We now turn to the description of the class of functions in which
we shall seek localized asymptotic solutions of the equation
(\ref{tr1.1})-(\ref{phaz_lst1.3a}). The functions of this class,
which singularly depend on a small parameter $\hbar$, represent a
generalized notion of a solitary wave. They depend on an arbitrary
phase trajectory $z=Z(t,\hbar)\in{\mathbb R}^{2n}_{p,x}$,
$t\in{\mathbb R}^1$ and on a real-valued function $S(t,\hbar)$
(analog of a classical action for a linear case with
$\varkappa=0$). In the limit $\hbar\to0$, the functions of this
class are concentrated in a neighborhood of a point moving along a
given phase curve $z=Z(t,0)$.  Functions of this type are well
known in quantum mechanics. In particular, among these are the
coherent and ``squeezed'' states of quantum systems with a
quadratic Hamiltonian \cite{Chern, Manko}.

Let us denote this class of functions by ${\mathcal
P}_\hbar^t(Z(t,\hbar),S(t,\hbar))$ and define it as follows:
\begin{align}
{\mathcal P}_\hbar^t&={\mathcal P}_\hbar^t
  \big(Z(t,\hbar),S(t,\hbar)\big)=\nonumber\\
&=\biggl\{\Phi :\Phi (\vec
x,t,\hbar)=\varphi\Bigl(\frac{\Delta\vec
  x}{\sqrt{\hbar}},t,\sqrt\hbar\Bigr) \exp\Bigl[{\frac{i}{\hbar}(S(t,\hbar)+
  \langle \vec P(t,\hbar),\Delta\vec x\rangle)}\Bigr]\biggr\},
\label{phaz_lst1.1}
\end{align}
where the complex-valued function $\varphi(\vec \xi,t,\sqrt\hbar)$
belongs to the Schwarz space $\mathbb S$ in the variable $\vec
\xi\in{\mathbb R}^n$, smoothly depends on $t$ and regularly
depends on $\sqrt\hbar$ as $\hbar\to0$. Here, $\Delta\vec x=\vec
x-\vec X(t,\hbar)$, and the real function $S(t,\hbar)$ and
$2n$-dimensional vector function $Z(t,\hbar)=(\vec P(t,\hbar),\vec
X(t,\hbar))$, which characterize the class ${\mathcal
P}_\hbar^t(Z(t,\hbar),S(t,\hbar))$, regularly depend on
$\sqrt\hbar$ in a neighborhood of $\hbar=0$, such that
$S(0,\hbar)=0$, $Z(0,\hbar)=z_0=(\vec p_0,\vec x_0)$ is an
arbitrary point of the phase space ${\mathbb R}^{2n}_{px}$, and
{\em are to be determined} for $t>0$.

The functions belonging to the class ${\mathcal P}_\hbar^t$, at
any fixed point in time $t\in{\mathbb R}^1$, {\em are
concentrated}, in the limit $\hbar\to0$, in a neighborhood of a
point lying on the phase curve $z=Z(t,0)$, $t\in{\mathbb R}^1$
(the exact sense of this property is specified below by formulas
(\ref{phaz_lst1.6}), (\ref{phaz_lst1.7})). Therefore, it is
natural to give the functions of the class ${\mathcal P}_\hbar^t$
the name {\em trajectory-concentrated functions} as $\hbar\to0$.
The definition of the class of trajectory-concentrated functions
involves a phase trajectory $Z(t,\hbar)$ and a scalar function
$S(t,\hbar)$ as ``free'' parameters. It turned out that these
``parameters'' can be uniquely determined by the
Hamilton--Ehrenfest system (see Sect. 3), which corresponds to the
nonlinear ($\varkappa\ne0$) Hamiltonian of equation
(\ref{tr1.1})-(\ref{phaz_lst1.3a}). Note that for a Schr\"odinger
type equation, in the linear case ($\varkappa=0$), the vector
function $Z(t,0)$ --- the principal term of the expansion in
$\hbar\to0$ --- determines the phase trajectory of a Hamilton
system with a classical Hamiltonian ${\mathcal H}(\vec p,\vec
x,t)$, and the function $S(t,0)$ is a classical action along this
trajectory. In particular, in this case, the class ${\mathcal
P}_\hbar^t$ involves the well-known dynamic (squeezed) states of
quantum systems with quadratic Hamiltonians.

Let us consider the basic properties of functions of the class
${\mathcal P}_\hbar^t(Z(t,\hbar), S(t,\hbar))$ (see
{\rm\cite{shapovalov:BAGRE,shapovalov:BTS1,shapovalov:BTS2}}).

\begin{prop} \label{teor1}
For functions of the class ${\mathcal
P}_\hbar^t(Z(t,\hbar),S(t,\hbar))$, the following asymptotic
estimates are valid for the centered moments
$\Delta_\alpha(t,\hbar)$ of order $|\alpha|$, $\alpha\in{\mathbb
Z}_+^{2n}$:
\begin{gather}
\Delta_\alpha(t,\hbar)=\Delta_\alpha^\Phi(t,\hbar)=
  \frac{\langle\Phi|\{\Delta{\hat z}\}^\alpha|\Phi\rangle}
  {\|\Phi\|^2}=O(\hbar^{|\alpha|/2}),\quad \hbar\to0, \quad
  |\alpha|\ne0;\label{phaz_lst1.2}\\
\Delta_\alpha(t,\hbar)=1, \quad |\alpha|=0.\nonumber
\end{gather}
Here $\{\Delta{\hat z}\}^\alpha$ denotes an operator with the Weyl
symbol $(\Delta z)^\alpha$,
\[\Delta z=z-Z(t,\hbar)=(\Delta\vec p,\Delta\vec x),\qquad
  \Delta\vec p=\vec p-\vec P(t,\hbar),
  \qquad \Delta\vec x=\vec x-\vec X(t,\hbar). \]
\end{prop}

Denote by the symbol $\widehat O(\hbar^k)$, $k\ge0$, an operator
$\widehat F$, such that for any function $\Phi\in{\mathcal
P}_\hbar^t(Z(t,\hbar),S(t,\hbar))$ the following asymptotic
estimate is valid:
\begin{equation}
\frac{\|\widehat F\Phi\|}{\|\Phi\|}=O(\hbar^k), \qquad
\hbar\to0.\label{phaz_lst1.3}
\end{equation}

\begin{prop} \label{teor2}
For functions of the class ${\mathcal
P}_\hbar^t(Z(t,\hbar),S(t,\hbar))$, the following asymptotic
estimates are valid:
\begin{eqnarray}
& \{ -i\hbar\partial_t -\dot S(t,\hbar)+\langle\vec P(t,\hbar),
  \dot{\vec X}(t,\hbar)\rangle+\langle\dot Z(t,\hbar),J\Delta{\hat z}\rangle\}=
  \widehat O({\hbar}),\label{phaz_lst1.4}\\
& \{\Delta{\hat z}\}^\alpha =\widehat
O({\hbar}^{|\alpha|/2}),\quad \alpha\in{\mathbb Z}^{2n}_+,\quad
\hbar\to 0,\nonumber
\end{eqnarray}
and, in particular,
\begin{equation}
\Delta{\hat x}_k =\widehat O(\sqrt{\hbar}),\quad \Delta{\hat p}_j
=\widehat O(\sqrt{\hbar}), \quad
k,j=\overline{1,n}.\label{phaz_lst1.5}
\end{equation}
\end{prop}

The following property gives an exact sense to the notion of the
concentration, as $\hbar\to0$, in a neighborhood of a point on the
phase trajectory for functions of the class ${\mathcal P}^t_h$.

\begin{prop} \label{teor3}
For any function $\Phi( \vec x,t,\hbar)\in{\mathcal
P}_\hbar^t(Z(t,\hbar),S(t,\hbar))$, the following limiting
relations are valid:
\begin{eqnarray}
&&\displaystyle\lim_{\hbar\to 0}\frac1{\|\Phi\|^2}|\Phi(\vec
x,t,\hbar)|^2=\delta(\vec x-\vec X(t,0)),\label{phaz_lst1.6}\\
&&\displaystyle\lim_{\hbar\to 0} \frac 1{\|\tilde\Phi\|^2}|
\tilde\Phi(\vec p,t,\hbar)|^2=\delta(\vec p-\vec
P(t,0)),\label{phaz_lst1.7}
\end{eqnarray}
where $\tilde\Phi(\vec p,t,\hbar)=F_{\hbar,\vec x\to \vec p}\Phi(
\vec x,t,\hbar)$, $F_{\hbar,\vec x\to\vec p}$ is a Fourier
$\hbar$-transform {\rm\cite{shapovalov:KARASEVMASLOV}}.
\end{prop}

\begin{prop} \label{teor4}
Denote by $\langle\hat L(t)\rangle$ the mean value of an operator
$\hat L(t)$, $t\in{\mathbb R}^1$ selfadjoint in $L_2({\mathbb
R}^n_x)$ that is calculated by the function $\Phi( \vec
x,t,\hbar)\in{\mathcal P}_\hbar^t$. Then, for any function $\Phi(
\vec x,t,\hbar)\in{\mathcal P}_\hbar^t(Z(t,\hbar),S(t,\hbar))$ and
any operator $\hat A(t,\hbar)$ whose Weyl symbol $A(z, t,\hbar)$
satisfies Assumption \ref{assum1}, the following equality is
valid:
\begin{equation}
\lim_{\hbar\to 0} \langle\hat A(t,\hbar)\rangle=\lim_{\hbar\to 0}
\frac 1{\|\Phi\|^2} \langle\Phi(\vec x,t,\hbar)|\hat
A(t,\hbar)|\Phi(\vec x,t,\hbar)\rangle
=A(Z(t,0),t,0).\label{phaz_lst1.8}
\end{equation}
\end{prop}

By analogy with the linear theory \cite{shapovalov:BAGRE}, we give
the following definition.

\begin{defin}\label{def1} A solution $\Phi(\vec x,t,\hbar)$ of equation
\eqref{tr1.1}--\eqref{phaz_lst1.3a} in the class of functions
${\mathcal P}_\hbar^t$ is called semiclassically concentrated as
$\hbar\to0$ on the phase trajectory $Z(t,\hbar)$.
\end{defin}

The limiting character of the conditions (\ref{phaz_lst1.6}),
(\ref{phaz_lst1.7}) and the asymptotic character of the estimates
(\ref{phaz_lst1.2})--(\ref{phaz_lst1.5}), valid in the class of
trajectory-concentrated functions, make it possible to construct
the semiclassically concentrated solutions of the nonlocal
Gross--Pitaevsky equation  not exactly, but approximately. In this
case, the $L_2$-norm of the error is of order $\hbar^\alpha$,
$\alpha>1$, as $\hbar\to0$ on any finite time interval $[0,T]$.

Denote such an approximate solution by $\Psi_{\rm as}=\Psi_{\rm
as}(x,t,\hbar)$. It satisfies the following problem:
\begin{gather}
\Bigl[-i\hbar\frac\partial{\partial t}+\widehat{\mathcal H}(t)+
  \varkappa\widehat V(t,\Psi_{\rm as})\Bigr]\Psi_{\rm as}=O(\hbar^\alpha), \\
\Psi_{\rm as}\in{\mathcal P}_\hbar^t(Z(t,\hbar),S(t,\hbar),\hbar),
\quad
  t\in[0,T],\label{phaz_lst1.10}
\end{gather}
where $O(\hbar^\alpha)$ denotes the function $g^{(\alpha)}(\vec
x,t,\hbar)$ with represents the error for equation
(\ref{tr1.1})--(\ref{phaz_lst1.3a}) and obeys the estimate in the
norm $L_2$:
\begin{equation}
\max_{0\le t\le T} \|g^{(\alpha)}(\vec
x,t,\hbar)\|=O(\hbar^\alpha), \qquad
\hbar\to0.\label{phaz_lst1.11}
\end{equation}
We shall also call the function $\Psi_{\rm as}(\vec x,t,\hbar)$,
satisfying the problem (\ref{phaz_lst1.9})--(\ref{phaz_lst1.11}),
{\em a semiclassically concentrated solution} ($\bmod
\hbar^\alpha$, $\hbar\to0$) of the nonlocal Gross--Pitaevsky
equation (\ref{tr1.1})--(\ref{phaz_lst1.3a}).

Thus, the semiclassically concentrated solutions $\Psi_{\rm
as}(\vec x,t,\hbar)$ of the nonlocal Gross--Pitaevsky equation
approximately describe the evolution of the initial state
$\psi_0(\vec x,\hbar)$ if it is chosen in the class of the
trajectory-concentrated functions ${\mathcal P}_\hbar^0$:
\begin{equation}
\Psi(\vec x,t,\hbar)|_{t=0}=\psi_0(\vec x,\hbar), \qquad \psi_0\in
{\mathcal P}^0_\hbar(z_0,0).\label{phaz_lst1.12}
\end{equation}
The functions of the class ${\mathcal P}_\hbar^0$ have the form
\begin{equation}
\psi_0(\vec x,\hbar)=\exp\Bigl\{\frac{i}\hbar[\langle \vec p_0,
\vec x-\vec x_0\rangle]\Bigr\}\varphi_0\Bigl(\frac{\vec x-\vec
x_0} {\sqrt\hbar},\sqrt\hbar\Bigr),\quad \varphi_0(\vec
\xi,\sqrt\hbar)\in{\mathbb S}({\mathbb R}^n_\xi),
\label{phaz_lst1.13}
\end{equation}
where $z_0=(\vec p_0,\vec x_0)$ is an arbitrary point of the phase
space ${\mathbb R}^{2n}_{px}$.

As in the linear case \cite{vs2}, among the solutions of equation
(\ref{tr1.1}) that satisfy the quasi-periodicity condition
(\ref{tr1.2aa}), in the class of solutions concentrated in a
neighborhood of some phase trajectory, we can isolate a set of
semiclassical asymptotics $\Psi_{{\mathcal E}_\nu}$ which possess
the following properties:

1) $\Psi_{{\mathcal E}_\nu}(\vec x,t,\hbar)$ are approximate in
$\bmod\,O(\hbar^{5/2})$ solutions of equation (\ref{tr1.1}). This
implies that
\begin{equation}
\begin{gathered}
\big[-i\hbar\partial_t + \widehat{\mathcal H}(t)+
  \varkappa\widehat V(t,\Psi(t))\big]\Psi_{{\mathcal E}_\nu}(\vec
  x,t,\hbar)=v_\nu(\vec x,t,\hbar),\\
\max_{t\in [0,T]}\| v_\nu(\vec x,t,\hbar)\|_{L_2({\mathbb
  R}_q^n)}=O(\hbar^{5/2}).\end{gathered} \label{phaz_lst1.9}
\end{equation}

2) The functions $\Psi_{{\mathcal E}_\nu}$ have the form of wave
packets concentrated at any time $t$ in a neighborhood of a given
$T$-periodic trajectory $z=Z(t,\hbar)$.  We shall call these
states, if any, the quasi-energy trajectory-coherent states
(TCS's) of the nonlocal Gross--Pitaevsky equation.

\section{System of Hamilton--Ehrenfest equations}

The symbols ${\mathcal H}(z,t)$ and $V(z,w,t)$ satisfy the
conditions of Assumption \ref{assum1}. Therefore, the operator
${\mathcal H}(\hat z,t)$ in (\ref{tr1.1})--(\ref{phaz_lst1.3a}) is
selfadjoint with respect to the scalar product
$\langle\Psi|\Phi\rangle$ in the space $L_2({\mathbb R}_x^n)$ and
the operator $V(\hat z,\hat w,t)$ with respect to the scalar
product $L_2({\mathbb R}^{2n}_{xy})$:
$\langle\Psi(t)|\Phi(t)\rangle_{{\mathbb R}^{2n}}
=\int\limits_{{\mathbb R}^{2n}} d\vec xd\vec y {\Psi}^* (\vec
x,\vec y,t,\hbar) \Phi(\vec x,\vec y,t,\hbar).$ Hence, the squared
norm of the exact solutions of equation
(\ref{tr1.1})--(\ref{phaz_lst1.3a}) is reserved: $\|\Psi(t)\|^2
=\|\Psi(0)\|^2$, and for the mean values of the operator $\hat
A(t)=A(\hat z,t)$ calculated for these solutions
\begin{equation}
\frac{d}{dt}\langle\hat A(t)\rangle
=\Bigl\langle\frac{\partial\hat A(t)} {\partial t}\Bigr\rangle
+\frac{i}{\hbar}\langle[\hat {\mathcal H},\hat A (t)]\rangle+
\frac{i\varkappa}{\hbar}\Bigl\langle\int d\vec y\,\Psi^*(\vec
y,t,\hbar) [V(\hat z,\hat w,t),\hat A (t)]\Psi(\vec
y,t,\hbar)\Bigl\rangle, \label{phaz_lst2.1}
\end{equation}
where $[\hat A,\hat B]=\hat A\hat B-\hat B\hat A$ is the
commutator of the operators $\hat A$ and $\hat B$. By analogy with
a linear ($\varkappa=0$) Schr\"odinger equation in quantum
mechanics, we call equality (\ref{phaz_lst2.1}) {\em the Ehrenfest
equation for the mean values of the operator} $\hat A(t)$ which
corresponds to the nonlocal Gross--Pitaevsky equation.


We suppose that for the nonlocal Gross--Pitaevsky equation
(\ref{tr1.1})--(\ref{phaz_lst1.3a}) there exist exact (or
differing by $O(\hbar^\infty)$ from exact) solutions in the class
of trajectory-concentrated functions. Let us write the Ehrenfest
equations (\ref{phaz_lst2.1}) for the mean values of the operators
$\hat z$, $\{\Delta\hat z\}^\alpha$, which are calculated by this
type (trajectory-concentrated) solutions of equation
(\ref{tr1.1})--(\ref{phaz_lst1.3a}), with the use of the
composition rules for the Weyl symbols
\cite{shapovalov:KARASEVMASLOV}:
\[C(z)=A\Big(\stackrel{\scriptscriptstyle2}{z}+\frac{i\hbar}2J
  \frac{\stackrel{\scriptscriptstyle1}{\partial}}{\partial z}\Big)
  B(z)=B\Big(\stackrel{\scriptscriptstyle2}{z}-\frac{i\hbar}2J
  \frac{\stackrel{\scriptscriptstyle1}{\partial}}{\partial z}\Big) A(z),\]
where $C(z)$ is the symbol of the operator $\hat C=$ $\hat A\,\hat
B$ and the numbers 1 and 2 above an operator indicate the order of
its action (recall that $\hat z =(\hat p,\hat x)$, $Z(t,\hbar)=
(\vec P(t,\hbar),\vec X(t,\hbar))$, $\Delta\hat z=\hat
z-Z(t,\hbar))$. Then, after calculations similar to that in the
linear case $\varkappa=0$ (see for details \cite{shapovalov:BAGRE,
88a}), restricting ourselves to the second-order moments, we
obtain the following system of ordinary differential equations:
\begin{equation}
\begin{aligned}
&\dot z=J\partial_z\Bigl(1+\frac12\langle\partial_z,
 \Delta_2\partial_z\rangle+\frac 12\langle\partial_w,
 \Delta_2\partial_w\rangle\Bigl)({\mathcal H}(z,t)+
 \tilde{\varkappa}V(z,w,t)|_{w=z}),\\
&\dot\Delta_2=J{\mathfrak H}_{zz}(z,t)\Delta_2-\Delta_2 {\mathfrak
H}_{zz}(z,t)J,
\end{aligned}\label{phaz_lst2.7}
\end{equation}
where $\Delta_2=(\Delta^{\alpha_1\alpha_2}_2)$ is a $(2n\times2n)$
 matrix of ``variances'', the $(2n\times2n)$ matrix
${\mathfrak H}_{zz}(z,t)$ is defined by
$$
{\mathfrak H}_{zz}(z,t)=\bigl[{\mathcal
H}_{zz}(z,t)+\tilde\varkappa V_{zz}(z,w,t) \Bigl|_{w=z}\bigr].
$$
and $\tilde{\varkappa}=\varkappa\|\psi_0(\vec x,\hbar)\|^2$,
$\psi_0(\vec x,\hbar)$ denotes the initial condition of the Cauchy
problem (\ref{phaz_lst1.12}).

System (\ref{phaz_lst2.7}) can be written in an equivalent,
setting in the second equation
\[\Delta_2(t)= A(t)\Delta_2(0) A^+(t), \]
which then takes the form
\begin{equation}
\dot A=J{\mathfrak H}_{zz}(z,t)A \qquad A(0)=\mathbb I.
\label{lkst2.6a}
\end{equation}
We shall call the system of equations (\ref{phaz_lst2.7}),
(\ref{lkst2.6a}) the second order {\em Hamilton--Ehrenfest} (HE)
{\em system} (here $M=2$ is the order of the greatest moment taken
into account) corresponding to the Gross--Pitaevsky
equation(\ref{tr1.1})--(\ref{phaz_lst1.3a}).

\begin{teo} Let functions $\Psi^{(2)}(\vec x,t,\hbar)$ and $\Psi(\vec x,t,\hbar)$
be, respectively, an asymptotic, to within $O(\hbar^{3/2})$, and
an exact solution of equation \eqref{phaz_lst1.1}, which coincide
at some point in time $t=t_0$. Then the second order
Hamilton--Ehrenfest systems \eqref{phaz_lst2.7}, constructed for
the functions $\Psi^{(2)}(\vec x,t,\hbar)$ and $\Psi(\vec
x,t,\hbar)$, coincide.
\end{teo}

\begin{teo}
Let functions $\Psi^{(2)}(\vec x,t,\hbar)$ and $\Psi(\vec
x,t,\hbar)$ be, respectively, an asymptotic, to within
$O(\hbar^{3/2})$, and an exact solution of equation
\eqref{phaz_lst1.1}, which coincide at some point in time $t=t_0$.
Then
\begin{equation}
\begin{gathered}
\Delta_{2\Psi}(t,\hbar)=\Delta_{2\Psi^{(2)}}(t,\hbar)+O(\hbar^{3/2})=
 \Delta_2(t,\hbar)+O(\hbar^{3/2}), \\
z_{\Psi}(t,\hbar)=z_{\Psi^{(2)}}(t,\hbar)+O(\hbar^{3/2})=
 Z^{2}(t,\hbar)+O(\hbar^{3/2}).\end{gathered} \label{phaz_lst2.8}
\end{equation}
\end{teo}

Let us consider the HE system (\ref{phaz_lst2.7}) as an abstract
system of ordinary differential equations with arbitrary initial
conditions. Obviously, not all solutions of the HE system
(\ref{phaz_lst2.7}) can be obtained by averaging the corresponding
operators over the solutions of the Gross--Pitaevsky equation
(\ref{tr1.1}). For instance, the mean values should satisfy the
Schr\"odinger--Robertson uncertainty relation \cite{Robertson},
which, for the second order moments 
implies nonnegative determinacy of the matrix
\begin{equation}
\Delta_2(t)-\frac{i\hbar}2 J.\label{a27}
\end{equation}
In the one-dimensional case, condition (\ref{a27}) is equivalent to
the well-known Schr\"odinger  uncertainty relation \cite{DoMa1}
\begin{equation}
\sigma_{pp}\sigma_{xx}-\sigma_{xp}^2\ge\frac{\hbar^2}4.\label{lst04a4a}
\end{equation}

Denote by ${\mathfrak g}(t,{\mathfrak C})$ the general solution of
the system of Hamilton--Ehrenfest equations (\ref{phaz_lst2.7}):
\begin{equation}
{\mathfrak g}(t,{\mathfrak C})=\big(\vec P(t,\hbar,{\mathfrak
C}),\vec X(t,\hbar,{\mathfrak C}),\Delta_{11}(t,\hbar,{\mathfrak
C}), \Delta_{12}(t,\hbar,{\mathfrak C}), \ldots , \Delta_{2n
2n}(t,\hbar,{\mathfrak C})\big)^\intercal
\label{shapovalov:HES-SOL}
\end{equation}
and by $\hat{\mathfrak g}$ the operator column
\begin{equation}
\hat{\mathfrak g}=\big(\hat{\vec p}, \vec{ x},(\Delta\hat
p_1)^2,\Delta\hat p_1\Delta\hat p_2,\ldots,(\Delta
x_n)^2\big)^\intercal.\label{shapovalov:HES-G}
\end{equation}
Here,
\begin{equation}\label{shapovalov:CONS-0}
{\mathfrak C}=(C_1,\ldots,C_N)^\intercal\in {\mathbb R}^{3n+2n^2}
\end{equation}
are arbitrary constants, and $B^\intercal$ denotes the transpose
to the matrix $B$.

These constants ${\mathfrak C}$,  Eq. (\ref{shapovalov:HES-SOL})
specify the trajectory of a point in the phase space ${\mathcal
M}^N$.

\begin{lem}\label{shapovalov:THEOR1}
Let $\Psi(\vec x,t)$ be a particular solution of equation {\rm
(\ref{phaz_lst1.1})} with the initial condition  $\Psi(\vec
x,t)\big|_{t=0}= \psi(\vec x)$. Determine the constants
${\mathfrak C}(\Psi(t))$ from the system
\begin{equation}
{\mathfrak g}(t,{\mathfrak C})=\frac{1}{\|\Psi \|^2} \langle
\Psi(t)|\hat{\mathfrak g}| \Psi(t)\rangle+O(\hbar^{3/2}),
\label{shapovalov:CONS-1}
\end{equation}
and the constants ${\mathfrak C}(\psi)$ from the system
\begin{equation}
{\mathfrak g}(0,{\mathfrak C})=\frac{1}{\|\psi\|^2}
\langle\psi|\hat{\mathfrak g}|\psi\rangle+
O(\hbar^{3/2}).\label{shapovalov:CONS-2}
\end{equation}
Then
\begin{equation}
{\mathfrak C}(\Psi(t))={\mathfrak C}(\psi)+O(\hbar^{3/2});
\label{shapovalov:CONS-2a}
\end{equation}
that is, ${\mathfrak C}(\Psi(t))$ are asymptotic, to within
($O(\hbar^{3/2})$), integrals of motion for equation
\eqref{phaz_lst1.1}.
\end{lem}
{\bf Proof}. By construction, the vector
\begin{equation}
{\mathfrak g}(t)=\frac{1}{\|\Psi\|^2}\langle\Psi(t)|
\hat{\mathfrak g}|\Psi(t)\rangle={\mathfrak g}(t,{\mathfrak
C}(\Psi(t)))\label{shapovalov:CONS-3}
\end{equation}
is a particular solution (to within $O(\hbar^{3/2})$) of system
(\ref{phaz_lst2.7}), which coincides with ${\mathfrak
g}(t,{\mathfrak C}(\psi))$ at $t=0$. By virtue of the uniqueness
of the Cauchy problem for system (\ref{phaz_lst2.7}), the relation
\begin{equation}
{\mathfrak g}(t,{\mathfrak C}(\psi))=
 {\mathfrak g}(t,{\mathfrak C}(\Psi(t)))+O(\hbar^{3/2})\label{shapovalov:CONS-4}
\end{equation}
holds true.


By virtue of the estimates (\ref{phaz_lst1.2}), the
Hamilton--Ehrenfest system can be solved not exactly, but
approximately:
\begin{equation}
\begin{array}{l}Z(t)=Z_0(t)+\hbar Z_1(t)+O(\hbar^2),\\[6pt]
\Delta_2(t)=\Delta_2^{(0)}(t)+O(\hbar^2).\end{array}\label{kvs7}
\end{equation}
Substituting (\ref{kvs7}) to system (\ref{phaz_lst2.7}), we
obtain, to within  $O(\hbar^{3/2})$, a system of equation for
$z_0=Z_0$, $z_1=Z_1$, $\Delta_2=\Delta_2^{(0)}(t)$
\begin{equation}
\begin{cases}{\dot z}_0=J\partial_{z_0}[{\mathcal H}(z_0,t) +\tilde\varkappa
  V(z_0,w,t)]\Big|_{w=z_0}, \\
{\dot z}_1=J[{\mathfrak H}_{zz}(z_0,t)+\tilde\varkappa V_{zw}(z_0,z_0,t)]z_1 +\\
\qquad +\dfrac{1}{2\hbar}
J\partial_{z_0}\mbox{Sp}\bigl\{[{\mathcal
H}_{zz}(z_0,t)+\tilde\varkappa V_{zz}(z_0,w,t)+
\tilde\varkappa V_{ww}(z_0,w,t)]\Delta_2\bigr\}\Big|_{w=z_0},\\
\dot\Delta_2=J{\mathfrak H}_{zz}(z_0,t)\Delta_2-\Delta_2{\mathfrak
H}_{zz}(z_0,t)J.
\end{cases}\label{kvs8}
\end{equation}
The first equation of (\ref{kvs8}) is a generalization of the
system of Hamilton equation to the case of a self-action.

If a solution $z_0=Z_0(t)$ of this system is known and also known
is the set of solutions of two systems in variations (for this to
be the case, it is sufficient that the matrix $V_{zw}(z,z,t)$ be
symmetric)
\begin{equation}
\dot a_k=J{\mathfrak H}_{zz}(t)a_k, \quad k=\overline{1,n},\quad
{\mathfrak H}_{zz}(t)={\mathfrak H}_{zz}(Z_0(t),t) \label{kvs9}
\end{equation}
normalized by the condition
\begin{equation}
\{a_k(t),a_l(t)\}=\{a_k^*(t),a_l^*(t)\}=0, \quad
\{a_k(t),a_l^*(t)\}=2 i\delta_{kl},\label{kvs10}
\end{equation}
where $\{a,b\}$ is a skew-scalar product:
\begin{eqnarray}
&\{a,b\}=\langle Ja,b\rangle=\langle\vec W_a,\vec Z_b\rangle -
\langle\vec Z_a,\vec W_b\rangle,\label{kvs11}\\
& a=\begin{pmatrix}\vec W_a\\ \vec Z_a\end{pmatrix}, \quad
 b=\begin{pmatrix}\vec W_b\\ \vec Z_b\end{pmatrix}, \nonumber
\end{eqnarray}
and of the system
\begin{equation}
\dot{\mathfrak a}_k = J\widetilde{\mathfrak H}_{zz}(t){\mathfrak
a}_k, \quad k=\overline{1,n},\quad \widetilde{\mathfrak
H}_{zz}(t)={\mathfrak H}_{zz}(Z_0(t),t)+\tilde\varkappa
V_{zw}(Z_0(t),Z_0(t),t), \label{kvs9a}
\end{equation}
also normalized by (\ref{kvs10}). Then the general solution of the
two last equations of (\ref{kvs8}) has the form
\begin{align}
& Z_1(t)=\sum_{k=1}^n [b_k(t){\mathfrak a}_k(t)+b_k^*(t){\mathfrak a}_k^*(t)],\label{kvs12}\\
& \Delta_2(t)=A(t){\mathcal D} A^\intercal(t),\label{kvs12a}
\end{align}
where
\begin{align}
& b_k(t)=\frac1{2i}\int\limits_0^t\{F(t),{\mathfrak a}_k^*(t)\} dt
+
B_k,\nonumber\\
& F(t)=\frac{1}{2\hbar} J\partial_{z}\mbox{Sp}\bigl\{ [{\mathcal
H}_{zz}(z,t)+\tilde\varkappa V_{zz}(z,w,t)+ \tilde\varkappa
V_{ww}(z,w,t)]\Delta_2\Bigr\}\Big|_{w=z=Z_0(t)},\label{kvs11a}\\
& A(t)=\big(a_1(t),a_2(t),\ldots,a_n(t),a_1^*(t),
a_2^*(t),\ldots,a_n^*(t)\big).\nonumber
\end{align}
Here, $B_k$ are integration constants and $\mathcal D$ is an
arbitrary constant matrix. Thus, in this approximation, the
solution of the problem is completely determined by the solution
of the generalized Hamilton system and the system in variations.

The quasi-periodicity condition (\ref{tr1.2aa}) leads to the
following constraint on the solutions of the Hamilton--Ehrenfest
system:
\begin{equation}
Z_0(t+T) = Z_0(t), \quad Z_1(t+T) = Z_1(t), \quad
 \Delta_2(t+T)=\Delta_2(t) \label{tr2.2a}
\end{equation}
We shall denote by ${\mathfrak C}_T$ the values of the constants
${\mathfrak C}$ at which condition (\ref{tr2.2a}) is fulfilled.

In this case, the systems in variations (\ref{kvs9}),
(\ref{kvs9a}) are systems of ordinary differential equations with
periodic coefficients
\[ {\mathfrak H}_{zz}(t+T)={\mathfrak H}_{zz}(t),\quad
  \widetilde{\mathfrak H}_{zz}(t+T)=\widetilde{\mathfrak H}_{zz}(t). \]
State for the systems in variations (\ref{kvs9}), (\ref{kvs9a})
the Floquet problems:
\begin{equation}
a_k(t+T) = {\rm e}^{i\Omega_kT}a_k(t),\quad {\mathfrak a}_k(t+T) =
{\rm e}^{i\widetilde\Omega_kT}{\mathfrak a}_k(t).\label{tr2.3}
\end{equation}

Suppose that each Floquet problem of (\ref{kvs9}), (\ref{kvs9a}),
(\ref{tr2.3}) has $n$, $n=\dim{\mathbb R}^n_x$, linearly
independent Floquet solutions $a_k(t) = (\vec W_k(t),\vec
Z_k(t))^\intercal$, ${\mathfrak a}_k(t) = (\vec{\widetilde
W}_k(t),\vec {\widetilde Z}_k(t))^\intercal$, with pure imaginary
Floquet indices $i\Omega_n$,$i\widetilde\Omega_n$, which satisfy
the orthogonality and normalization condition (\ref{kvs10}).

It should be stressed that in the context of the Floquet theory
for linear Hamilton systems with periodic coefficients, the
conditions $\mbox{Im}\;\widetilde\Omega_k=0$ imply stability of
the phase trajectory $z=Z_0(t)$ in the linear approximation.
Recall that the $2n$-dimensional vectors $a_k(t)$ and $a^*_k(t)$,
$k=\overline{1,n}$, constitute a simplectic basis in ${\mathbb
C}^{2n}_a$, and the $n$-dimensional plane $r^n(Z(t,\hbar))$
spanned over the vectors $a_k(t)$ constitutes a complex germ on
$Z(t,\hbar)$ \cite{shapovalov:MAS,BeD2}.

Determine the integration constants $B_{k}$ from the condition of
periodicity of the function $\hbar Z_1(t)$ in time:
\begin{equation}
 Z_1(t+T)= Z_1(t). \label{phazcyc3.14}
\end{equation}
Like for (\ref{kvs12}), we put
\begin{equation}
b_k(t)=\frac1{2i}\int\limits_0^t d\tau \{F(\tau), {\mathfrak
a}^*_k(\tau) \}+B_k. \label{phazcyc3.15}
\end{equation}
Then conditions (\ref{tr2.3}), \eqref{phazcyc3.14} imply that
\begin{equation}
b_k(t+T)= b_k (t)e^{-i\widetilde\Omega_kT}. \label{phazcyc3.16}
\end{equation}
Note that
\begin{align*}
b_k(t+T)&=\frac1{2i}\int\limits_0^{t+T} d\tau \{F(\tau),
{\mathfrak a}^*_k(\tau)\}+B_k=b_k(T)+\frac1{2i}\int\limits_T^{t+T}
d\tau
\{F(\tau), {\mathfrak a}^*_k(\tau)\}={}\\
& =b_k(T)+\frac1{2i}\int\limits_0^t d\tau \{ F (\tau+T),
{\mathfrak a}^*_k(\tau+T)\}=b_k(T)+e^{-i\widetilde\Omega_kT}
(b_k(t)-B_k).
\end{align*}
Hence,
\begin{equation}
B_k=b_k(T)e^{i\widetilde\Omega_kT}.\label{phazcyc3.17}
\end{equation}

\section{The linear associated Schr\"odinger equation}

The linearization of the nonlocal Gross--Pitaevsky equation in the
class of trajectory-concentrated functions is the central point in
our approach.

Introduce the notation (see (\ref{phaz_lst1.2}))
\begin{equation}
{\mathfrak g}_\Psi(t)=\frac{1}{\|\Psi \|^2}\langle\Psi(t)|
\hat{\mathfrak g}| \Psi(t)\rangle
\label{phaz_lst3.1}
\end{equation}
and expand the ``kernel'' of the operator $\widehat V(t,\Psi)$ in
a Taylor series in powers of the operators $\Delta\hat w=\hat
w-z_\Psi(t,\hbar)$. Substituting this expansion in equations
(\ref{tr1.1})--(\ref{phaz_lst1.3a}), for the functions
$\Psi\in{\mathcal P}_\hbar^t$, in view of the estimate
(\ref{phaz_lst1.2}), we obtain
\begin{equation}
\hat L^{(2)}(t,\Psi)\Psi=\Bigl\{-i\hbar\partial_t+{\mathfrak
H}(\hat z,t)+\tilde\varkappa V(\hat z,w,t)+
\frac{\tilde\varkappa}{2}\mbox{Sp}[V_{ww}(\hat z,w,t)
\Delta_{2\Psi}(t,\hbar)]\Bigr\}\Big|_{w=z_\Psi(t,\hbar)}\Psi= \hat
O(\hbar^{3/2}).\label{phaz_lst3.4}
\end{equation}

Let us associate the nonlinear equation (\ref{phaz_lst3.4}) with the
linear equation that is obtained from (\ref{phaz_lst3.4}) by
substituting the corresponding solutions of the HE system
${\mathfrak g}(t,{\mathfrak C}_T)$ for ${\mathfrak g}_\Psi(t)$, the
mean values of the operators of coordinates, momenta, and second
order centered moments. As a result, we obtain the following
equation:
\begin{align}
& \hat L^{(2)}(t,{\mathfrak C}_T)\Phi=0, \quad \Phi\in{\mathcal
P}^t_\hbar,
\label{phaz_lst3.8}\\
&\hat L^{(2)}(t,{\mathfrak C}_T)=-i\hbar\partial_t+{\mathcal
H}(\hat z,t) +\tilde\varkappa
 V(\hat z, w,t) \Big|_{w=Z(t,{\mathfrak C}_T)}+
 \frac{\tilde\varkappa}{2}\mbox{Sp}\, \Big[V_{ww}(\hat z,
w,t) \Big|_{w=Z(t,{\mathfrak C}_T)}\Delta_{2}(t,{\mathfrak
C}_T)\Big]. \label{phaz_lst3.7}
\end{align}
By virtue of (\ref{tr2.2a}), equations (\ref{phaz_lst3.7}) are
equations with periodic coefficients. The quasi-periodicity
condition (\ref{tr1.2aa}) reduces, for equation
(\ref{phaz_lst3.7}), the Floquet problem
\begin{equation}
\Phi_{\mathcal E}(\vec x,t+T,\hbar)=e^{-i{\mathcal E}T/\hbar}
\Phi_{\mathcal E}(\vec x,t,\hbar).\label{tr1.2Phi}
\end{equation}

Thus, the change of the quantum means of the operators $\hat w$
and $\{\Delta\hat w\}^\alpha$ by the solutions of the second order
Hamilton--Ehrenfest system in equation (\ref{phaz_lst3.4}) {\em
linearizes} the nonlocal Gross--Pitaevsky equation
(\ref{tr1.1})--(\ref{phaz_lst1.3a}) to within $O(\hbar^{3/2})$.
Hence, to construct the semiclassically concentrated states
$\bmod\hbar^{3/2}$ of the nonlocal Gross-Pitaevsky equation
(\ref{tr1.1})--(\ref{phaz_lst1.3a}), it suffices to construct,
with the same accuracy, an asymptotics of the solution of a
Schr\"odinger type {\it linear equation}.

\begin{defin}
We call equations of the form \eqref{phaz_lst3.8} parametrized by
constants ${\mathfrak C}$ \eqref{phaz_lst1.12} the set of
associated linear Schr\"odinger equations for the nonlocal
Gross--Pitaevsky equation \eqref{tr1.1}--\eqref{phaz_lst1.3a}, and
denote by $\Phi=\Phi(\vec x,t,\hbar,{\mathfrak C})$ its
(arbitrary) solution in the class of functions ${\mathcal
P}_\hbar^t$ \footnote{thereby stressing the dependence of the
solution of equation (\ref{phaz_lst3.8}) (via the coefficients of
the operator $\hat L^{(2)}(t,{\mathfrak C})$) on the solution
${\mathfrak g}(t,\hbar,{\mathfrak C})$ of the Hamilton--Ehrenfest
equation (\ref{phaz_lst2.7}) }.
\end{defin}

The following statement is obvious.

\begin{demo}
\label{d1} If a function $\Phi^{(2)}(\vec x,t,\hbar,{\mathfrak
C})\in{\mathcal P}_\hbar^t$
--- an asymptotic $($to within $O(\hbar^{3/2})$, $\hbar\to0)$ solution of equation
\eqref{phaz_lst3.8} --- satisfies the initial condition
\eqref{tr1.2Phi}:
\begin{equation}
\Phi^{(2)}(\vec x,t,\hbar,{\mathfrak C}(\psi_0))|_{t=0}=\psi_0,
\label{phaz_lst3.9}
\end{equation}
then the function $\Psi^{(2)}(\vec x,t,\hbar)=\Phi^{(2)}(\vec
x,t,\hbar,{\mathfrak C}(\psi_0))$  is a asymptotic $($to within
$O(\hbar^{3/2})$, $\hbar\to0)$ solution of the Cauchy problem for
the nonlocal Gross--Pitaevsky equation \eqref{tr1.1}. Constants
${\mathfrak C}(\psi_0)$ are determined by the equation
\eqref{shapovalov:CONS-2}.
\end{demo}

To construct the function $\Phi^{(2)}(\vec
x,t,\hbar,\psi_0)\in{\mathcal P}_\hbar^t(Z(t,\hbar),S(t,\hbar))$,
we shall use the technique for constructing localized asymptotics
developed in \cite{shapovalov:MAS,BeD2,bbt82, bbt83,vs2} for
linear quantum-mechanics equations, modifying this technique
taking into account the initial statement of the problem
(\ref{phaz_lst1.9})--(\ref{phaz_lst1.12}) and the structure of the
Hamiltonian (\ref{phaz_lst3.7}) for the (linearized) nonlocal
Gross--Pitaevsky equation in the trajectory-coherent
approximation. Namely, we shall define the point of localization
in ${\mathbb R}_{p,x}^{2n}$ of the semiclassically concentrated
solution of equation (\ref{phaz_lst3.8}) --- ``parameter''
$Z(t,\hbar)$ entering into the definition of the class ${\mathcal
P}_\hbar^t$
--- as a natural projection of the solution
$y^2_{\psi_0}(t,\hbar)$ of the Hamilton--Ehrenfest system
(\ref{phaz_lst2.7}) onto the phase space, i.e. we set
$Z(t,\hbar)=Z^2(t,\hbar)=(\vec P^2(t,\hbar), \vec X^2(t,\hbar))$.
Define the function $S(t,\hbar)$ (second ``parameter'' of the
class ${\mathcal P}_\hbar^t$) as an analog of the classical action
along this $(z=Z^2(t,\hbar),t\in[0,T])$ phase trajectory by a
standard formula in a classical Hamiltonian corresponding not to
the main, but to the {\em total symbol }
\cite{shapovalov:KARASEVMASLOV} ${\mathcal H}_\varkappa^2(z,t)$ of
the quantum Hamiltonian in (\ref{phaz_lst3.7}), which, in view of
the estimates $\Delta_\alpha^2(t,\hbar)=O(\hbar^{|\alpha|/2})$,
has the form
\begin{equation}
{\mathcal H}_\varkappa^2(z,t)={\mathcal H}(z,t) +\tilde\varkappa
 V(z, w,t) \Big|_{w=Z(t,{\mathfrak C}_T)}+
 \frac{\tilde\varkappa}{2}\mbox{Sp}\Big[V_{ww}(z,w,t)\Big|_{w=Z(t,{\mathfrak C}_T)}
 \Delta_{2}(t,{\mathfrak C}_T)\Big]. \label{phaz_lst3.10}
\end{equation}
As a result, we have
\begin{equation}
S(t,\hbar)=S^{(2)}(t,\hbar)=\int\limits_0^t\bigl\{\langle \vec
P^{(2)}(\tau,\hbar),\dot{\vec X}{}^{(2)}(\tau,\hbar)\rangle-
{\mathcal
H}_\varkappa^{(2)}(Z^{(2)}(\tau,\hbar),\tau)\bigr\}d\tau.
\label{phaz_lst3.11}
\end{equation}
Introduce the notation
$$
S^{(2)}(\vec x,t,\hbar)=S^{(2)}(t,\hbar)+\langle \vec
P^{(2)}(t,\hbar),\vec x-\vec X^{(2)}(t,\hbar)\rangle.
$$
Expand the operators ${\mathcal H}(\hat z,t)$ and
$\dfrac{\partial^{|\alpha|}}{\partial w^\alpha}V(\hat
z,w,t)|_{w=Z^{(2)}(t,\hbar)}$ in Taylor series of order two in
powers of the operator $\Delta\hat z=\hat z-Z^{(2)}(t,\hbar)$ with
remainder terms $\hat R^H_{3}$ and $\hat R^V_{3}$, respectively,
and represent the operator $-i\hbar\partial/\partial t$ in the
form
\begin{eqnarray}
&& -i\hbar\partial_t=\hat A+\hat B, \cr && \hat B=-\langle \vec
P^{(2)}(t,\hbar),\dot X^{(2)}(t,\hbar)\rangle+\dot
S^{(2)}(t,\hbar)-
\langle\dot Z^{(2)}(t,\hbar),J\Delta\hat z\rangle,\label{phaz_lst3.13} \\
&& \hat A=-i\hbar\partial_t-\dot S^{(2)}(t,\hbar) +\langle \vec
P^{(2)}(t,\hbar),\dot X^{(2)}(t,\hbar)\rangle+\langle\dot
Z^{(2)}(t,\hbar),J\Delta\hat z\rangle=\widehat{O}(\hbar).
\nonumber
\end{eqnarray}
Substituting the resulting expressions in equation
(\ref{phaz_lst3.8}), in view of the estimates in $\hbar\to0$, for
the operators $\hat R^H_{3}$ and $\hat R^V_{3}$ applied to
functions of the class ${\mathcal P}_\hbar^t$, we obtain
\begin{equation}
\{ -i\hbar\partial_t + \hat{\mathfrak H}_0(t,{\mathfrak
C}_T)\}\Phi =O(\hbar^{3/2}),\label{phaz_lst3.14}
\end{equation}
where
\begin{align}
&\hat{\mathfrak H}_0(t,{\mathfrak C}_T))=-\dot S^{(2)}(t,\hbar)+
  \langle\vec P^{(2)}(t,\hbar), \dot{\vec X}^{(2)}(t,\hbar)\rangle +
  \langle\dot Z^{(2)}(t,\hbar),J\Delta\hat z\rangle + 
  \frac12 \langle\Delta\hat z,{\mathfrak H}_{zz}(t, {\mathfrak C}_T))\Delta\hat z\rangle,\label{phaz_lst3.15}\\ 
&{\mathfrak H}_{zz}(t,{\mathfrak C}_T))=[{\mathcal
H}_{zz}(z,t)+\tilde\varkappa
  V_{zz}(z,w,t)]\Big|_{z=w=Z^{(2)}(t,\hbar)}.\label{phaz_lst3.17}
\end{align}
This is a Schr\"odinger equation with the Hamiltonian quadratic in
the operators $\hat{p}$ and $\hat{x}$.

\section{Trajectory-coherent states \\ of the nonlocal Gross--Pitaevsky equation}

The solution of the linear Schr\"odinger equation with a quadratic
Hamiltonian is well known. For our purposes, it is convenient to
take for the basis of solutions of equation (\ref{phaz_lst3.14})
the semiclassical trajectory-coherent states (TCS's) of this
equation \cite{bbt82, bbt83}. By Statement \ref{d1}\, these states
are asymptotic $(\bmod\,\hbar^{3/2})$ solutions of the problem
(\ref{phaz_lst1.9})--(\ref{phaz_lst1.12}) if the function
$\psi_0(x,\hbar)$ (\ref{phaz_lst1.12}) coincides with the TCS at
the time zero. We shall also call these solutions {\it
trajectory-coherent states of the nonlocal Gross--Pitaevsky
equation} (\ref{tr1.1})--(\ref{phaz_lst1.3a}). We now give their
explicit form and some properties, which will be used below to
solve the problem (\ref{phaz_lst1.9})--(\ref{phaz_lst1.12}) with
an arbitrary initial condition of the class of functions
${\mathcal P}_\hbar^0$.

Let us use the ``momentum'' and ``coordinate'' components of the
solution of the Floquet problem (\ref{kvs9}), (\ref{tr2.3}) to
compose $n\times n$ matrices
\begin{eqnarray}
& a(t,{\mathfrak C}_T))=(\vec W(t,{\mathfrak C}_T)),\vec Z(t,
{\mathfrak C}_T)))^\intercal,\cr & B(t)=(\vec W_1(t),\vec
W_2(t),\dots,\vec W_n(t)), \qquad C(t)=(\vec Z_1(t),\vec
Z_2(t),\dots,\vec Z_n(t)).\label{phaz_lst4.6z}
\end{eqnarray}

The linear Hamilton system (\ref{kvs9}) is known \cite{Arnold2} to
conserve the standard simplectic structure $\omega=d\vec p\wedge
d\vec x$ of the phase space ${\mathbb R}^{2n}_{p,x}$, and, hence,
the skew-scalar product $\langle Ja,b\rangle=\omega(a,b)$ of any
two complex solutions $a(t)$, $b(t)\in{\mathbb C}^{2n}_{w,z}$ of
system (\ref{kvs9}) does not depend on time. (Here, ${\mathbb
C}^{2n}_{w,z}$ denotes the complexification ${\mathbb
R}^{2n}_{p,x}$ with complex coordinates $w\in{\mathbb C}^n$,
$z\in{\mathbb C}^n$, $a=(\vec w,\vec z)$.) From this reasoning, in
view of (\ref{kvs10}), we obtain
\begin{eqnarray}
& C^\intercal(t)B(t)-B^\intercal
C(t)=\big\|{\{a_i(t),a_j(t)\}}\big\|_{n\times n}=0,
\label{phaz_lst4.5}\\
& \dfrac{1}{2i}(C^+B(t)-B^+C(t))=\dfrac{1}{2i}
\big\|{\{a_i(t),a_j^*(t)\}}\big\|_{n\times n}={\mathbb I},
\label{phaz_lst4.6}
\end{eqnarray}

Conventional reasoning (see, e.g., \cite{shapovalov:BAGRE}), in
view of (\ref{phaz_lst4.6}), lead to the statement that the matrix
$C(t)$ is nondegenerate, $\det C(t)\neq0$, and the imaginary part
of the matrix
\begin{equation}
Q(t)=B(t)C^{-1}(t) \label{phaz_lst4.7}
\end{equation}
is positive for $t\in[0,T]$, $T>0$: $\mbox{Im}\,Q(t)>0$. Moreover,
by virtue of (\ref{phaz_lst4.5}), the matrix $Q(t)$ is symmetric,
$Q^\intercal(t)=Q(t)$.

Let us now fix the continuous branch of the root of $\det C(t)$,
$t\in[0,T]$, assuming, for instance, that \linebreak
$\mbox{Arg}\sqrt{\det C(0)}=0$, and determine the function
\begin{equation}
\Phi^{(2)}_0(\vec x,t,\hbar)=|0,t,{\mathfrak C}_T\rangle
=\dfrac{N_0(\hbar)}{\sqrt{\det C(t)}}
\exp\Big\{\frac{i}{\hbar}\Big[S^{(2)}(t,\hbar)+\langle \vec
P^{(2)}(t,\hbar),
\Delta \vec x\rangle 
+\frac{1}{2}\langle\Delta \vec x,Q(t)\Delta \vec x\rangle
\Big]\Big\}, \label{phaz_lst4.8}
\end{equation}
where $\Delta \vec x=\vec x-\vec X^{(2)}(t,\hbar)$, $(\vec
P^{(2)}(t,\hbar),\vec X^{(2)}(t,\hbar))=Z^{(2)}(t,\hbar)$;
$t\in[0,T]$ is the phase trajectory by virtue of the second order
Hamilton--Ehrenfest system (\ref{kvs9}); $S^{(2)}(t,\hbar)$ is
defined in (\ref{phaz_lst3.10}), and
$N_0(\hbar)=(\pi\hbar)^{-n/4}$ is a normalizing constant: $\langle
t,0|0,t\rangle=1$.

\begin{teo}\label{teor-4.1}
The function $\Phi_0^{(2)}(\vec x,t,\hbar)$ \eqref{phaz_lst4.8} is
an (exact) solution of the Cauchy problem for equation
\eqref{phaz_lst3.14} with an initial condition of the form
\begin{equation}
\Phi_0^{(2)}(\vec
x,t,\hbar)|_{t=0}=N_0(\hbar)e^{\frac{i}{\hbar}[\langle \vec
p_0,\vec x-\vec x_0\rangle +\frac{1}{2}\langle \vec x-\vec
x_0,B(0)C^{-1}(0)(\vec x-\vec x_0)\rangle ]}, \label{phaz_lst4.9}
\end{equation}
where $(\vec p_0,\vec x_0)\in{\mathbb R}^{2n}_{p,x}$; the
$(n\times n)$ complex matrices $B(t)$ and $C(t)$ are defined in
{\rm(\ref{phaz_lst4.6z})}.
\end{teo}
{\bf Proof.} We seek a solution of the linear equation
(\ref{phaz_lst3.15}) in the class of functions ${\mathcal
P}_\hbar^t(S^{(2)}(t,\hbar),Z^{(2)}(t,\hbar))$ in the form of a
Gaussian packet
\begin{equation}
\Phi(\vec x,t,\hbar)=\exp\Big\{\frac{i}{\hbar}
[S^{(2)}(t,\hbar)+\langle \vec P^{(2)}(t,\hbar),\Delta \vec
x\rangle ]\bigg\} \exp\bigg\{\frac{i}{\hbar}\frac{\langle \Delta
\vec x,Q(t)\Delta \vec x\rangle}{2}\bigg\}\varphi(t),
\label{phaz_lst4.10}
\end{equation}
where the complex $(n\times n)$ matrix $Q(t)$ is symmetric and
$\mbox{Im}\,Q(t)>0$. Substituting (\ref{phaz_lst4.10}) in
(\ref{phaz_lst3.14}) and equating the coefficients of the powers
of the operator $\Delta x^k$, $k=0,2$, to zero, we obtain a linear
equation in the function $\varphi(t)$ and a (matrix) Riccati
equation in the matrix $Q(t)$, respectively:
\begin{eqnarray}
& \dot\varphi+\dfrac{1}{2}\mbox{Sp}[{\mathfrak
H}_{px}(t,{\mathfrak C}_T))+
{\mathfrak H}_{pp}(t,{\mathfrak C}_T))Q(t)]\varphi=0,\label{phaz_lst4.11}\\
& \dot Q+{\mathfrak H}_{xx}(t, {\mathfrak C}_T)+Q(t){\mathfrak
H}_{px}(t, {\mathfrak C}_T)+ {\mathfrak H}_{xp}(t, {\mathfrak
C}_T)Q(t)+Q(t){\mathfrak H}_{pp}(t, {\mathfrak C}_T)Q(t)=0.
\label{phaz_lst4.12}
\end{eqnarray}
The ordinary change of variables $Q(t)=B(t)C^{-1}(t)$ (see, e.g.,
\cite{shapovalov:BAGRE}), provided that $Q(0)=B(0)C^{-1}(0)$,
Reduces the problem of constructing the desired complex solution
of the equation (\ref{phaz_lst4.12}) to the problem
\begin{equation}
\begin{pmatrix} \dot B \\ \dot C\end{pmatrix}=
J{\mathfrak H}_{zz}(t,{\mathfrak C}_T) \begin{pmatrix} B \\
C
\end{pmatrix}\Longleftrightarrow
\begin{array}{l} \dot B=-{\mathfrak H}_{xp}(t,{\mathfrak C}_T)B-{\mathfrak H}_{xx}(t,{\mathfrak C}_T)C, \\
\dot C={\mathfrak H}_{pp}(t,{\mathfrak C}_T)B+{\mathfrak
H}_{px}(t,{\mathfrak C}_T)C,\end{array}\label{phaz_lst4.1}
\end{equation}
which, by virtue of (\ref{phaz_lst4.6z}), is reduced to system
(\ref{kvs9}). By virtue of the second equation of
(\ref{phaz_lst4.1}), we have
$$\dot C=[{\mathfrak H}_{px}(t, {\mathfrak C}_T))+{\mathfrak H}_{pp}(t,
{\mathfrak C}_T))Q(t)]C,
$$
where $Q(t)$ is a solution of (\ref{phaz_lst4.12}). Thus, by
virtue of Liouville's lemma, we find
$$
\det C(t)=\exp\int\limits_0^t\mbox{Sp}[{\mathfrak H}_{px}(\tau,
{\mathfrak C}_T))+ {\mathfrak H}_{pp}(\tau, {\mathfrak
C}_T))Q(\tau)]d\tau,
$$
and, hence, by equation (\ref{phaz_lst4.11}), we have
$\varphi(t)=(\det C(t))^{-1/2}$.

Let us now construct the Fock basis of solutions for equation
(\ref{phaz_lst3.14}). To do this, we find for this equation the
symmetry operators $\hat a(t, {\mathfrak C}_T))$, linear in
operators $\Delta\hat z$, in the form $\hat a(t, {\mathfrak
C}_T))=N_a\langle b(t, {\mathfrak C}_T)),\Delta\hat z\rangle,$
where $N_a$ is a constant, $b(t)=b(t, {\mathfrak C}_T))$ is the
complex $2n$ vector to be determined. By the equation
$$
-i\hbar\frac{\partial\hat a(t)}{\partial t}+[\hat{\mathfrak
H}_0(t, {\mathfrak C}_T)),\hat a(t)]=0,
$$
which determines the operator $\hat a(t)$, in view of the explicit
form of $\hat{\mathfrak H}_0(t, {\mathfrak C}_T))$
(\ref{phaz_lst3.15}), we obtain
\begin{eqnarray*}
&-i\hbar\langle \dot b(t),\Delta\hat z\rangle +i\hbar\langle b(t),\dot Z^{(2)}(t,\hbar)\rangle+\\
&+\Bigl[\Bigl\{-\dot S^{(2)}(t,\hbar)+\langle \vec
P^{(2)}(t,\hbar), \dot{\vec X}^{(2)}(t,\hbar)\rangle+ \langle \dot
Z^{(2)}(t,\hbar),J\Delta{\hat z}\rangle +\dfrac12\langle \Delta
z,{\mathfrak H}_{zz}(t, {\mathfrak C}_T)) \Delta\hat z\rangle
\Bigr\},\langle b(t),\Delta\hat z\rangle\Bigr]=0.
\end{eqnarray*}
From this equation, by virtue of the commutation relations
$[\Delta\hat z_j,\Delta\hat z_k]=i\hbar J_{jk}$,
$j,k=\overline{1,2n}$, we find $-i\hbar\langle\dot b(t),\Delta\hat
z\rangle+i\hbar\langle\Delta\hat z,{\mathfrak
H}_{zz}(t)Jb(t)\rangle=0,$ and, hence, $\dot b={\mathfrak
H}_{zz}(t, {\mathfrak C}_T))Jb$. Using the notation $b(t)=-Ja(t)$,
we obtain for the vector $a(t)$ the system in variations
(\ref{kvs9}). Thus, the operator
\begin{equation}
\hat a(t)=\hat a(t, {\mathfrak C}_T))=N_a\langle b(t),\Delta\hat
z\rangle= N_a\langle a(t),J\Delta\hat z\rangle
\label{phaz_lst4.13}
\end{equation}
is the symmetry operator for equation (\ref{phaz_lst3.14}) if the
vector $a(t)=a(t, {\mathfrak C}_T))$ is a solution of the system
in variations (\ref{kvs9}). Let $\hat a(t)$ and $\hat b(t)$ be
symmetry operators corresponding to two solutions of the system in
variations, $a(t)$ and $b(t)$, respectively. It is easy to check
that
\begin{equation}
[\hat a(t),\hat b(t)]=i\hbar N_aN_b\{a(t),b(t)\}=i\hbar
N_aN_b\{a(0),b(0)\}, \label{phaz_lst4.13'}
\end{equation}
the last equation being a corollary of the Hamiltonian character
of system (\ref{kvs9}).

By formula (\ref{phaz_lst4.13}), associate the vectors $a_j^*(t)$
with the ``creation'' operators $\hat a_j^+(t)$ and the vectors
$a_j(t)$ with the ``annihilation'' operators $\hat a_j(t)$,
setting $N_a=(2\hbar)^{-1/2}$. Then, by virtue of formulas
(\ref{phaz_lst4.13'}),for the operators $\hat a_j^+(t)$, $\hat
a_j(t)$, $j=\overline{1,n}$, the following commutation relations,
canonical for boson operators, are valid:
\begin{equation}
[\hat a_j(t),\hat a_k(t)]=[\hat a_j^+(t), \hat a_k^+(t)]=0, \quad
[\hat a_j(t),\hat a_k^+(t)]=\delta_{jk}, \quad j,k=\overline{1,n}.
\label{phaz_lst4.14}
\end{equation}

\begin{demo}
The function $\Phi_0^{(2)}(x,t,\hbar)=|0,t, {\mathfrak
C}_T)\rangle$ (\ref{phaz_lst4.8}) is a ``vacuum''
trajectory-coherent state:
\begin{equation}
\hat a_j(t)|0,t, {\mathfrak C}_T)\rangle =0, \quad
j=\overline{1,n}. \label{phaz_lst4.15}
\end{equation}
\end{demo}
{\bf Proof.} Applying the annihilation operator $\hat a_j(t)$ to
the function $|0,t\rangle$, we obtain that $\hat
a_j(0,t)|0,t\rangle=|0,t\rangle [\langle \vec Z_j(t),Q(t)\Delta
\vec x\rangle -\langle \vec W_j(t),\Delta x\rangle ]$. From this
(\ref{phaz_lst4.15}) immediately follows since, by definition and
in view of the properties of the matrix $Q(t)$, we have $Q(t)\vec
Z_j(t)=B(t)C^{-1}(t)\vec Z_j(t)=\vec W_j(t)$.

Let us now define a countable set of states $|\nu,t, {\mathfrak
C}_T)\rangle$ (exact solutions of equation (\ref{phaz_lst3.14}))
as the result of the action of the birth operators on the vacuum
state $|0,t, {\mathfrak C}_T)\rangle$ (\ref{phaz_lst4.8}):
\begin{equation}
\Phi^{(2)}_\nu(\vec x,t,\hbar)=|\nu,t,{\mathfrak
C}_T)\rangle=\frac 1{\nu!}(\hat a^+(t, {\mathfrak C}_T)))^\nu
|0,t, {\mathfrak C}_T)\rangle=\prod_{k=1}^n\frac 1{\nu_k!}(\hat
a_k^+(t, {\mathfrak C}_T)))^{\nu_k} |0,t, {\mathfrak C}_T)\rangle.
\label{phaz_lst4.16}
\end{equation}
The functions $\Phi_\nu^{(2)}(\vec x,t,\hbar)$, $\nu\in{\mathbb
Z}^n_+$ constitute the Fock basis of solutions of the (linear)
equation (\ref{phaz_lst3.14}): by formulas (\ref{phaz_lst4.15}),
(\ref{phaz_lst4.14}), standard calculations check whether this set
of functions is orthonormalized,
$\langle\Phi_\nu^{(2)}|\Phi_{\nu'}^{(2)}\rangle
=\delta_{\nu\nu'}$, $\nu,\nu'\in{\mathbb Z}_+^n$, and the proof of
its completeness follows, for example, from the results presented
in \cite{Simon}. Thus, from this reasoning and Statement
\ref{d1}\, we arrive at the following theorem.

\begin{teo} Let the symbols of the operators $\widehat{\mathcal H}(t)$ and
$\widehat V(t,\Psi)$ in \eqref{tr1.1}--\eqref{phaz_lst1.3a}
satisfy the conditions of Assumption \ref{assum1}\ and the
conditions of Theorem \ref{teor-4.1} be fulfilled. Then, for any
$\nu\in{\mathbb Z}^n_+$ the function
\begin{equation}
\label{shapovalov:CAUCHY-SOL} \Psi_\nu(\vec x,t,\hbar)=\Phi_\nu
(\vec x,t,\hbar,{\mathfrak C}^\nu_T)
\end{equation}
is an asymptotic $($to within $O(\hbar^{3/2})$, $\hbar\to0)$
solution of the nonlocal Gross--Pitaevsky equation
\eqref{tr1.1}--\eqref{phaz_lst1.3a} with a quasiperiodicity
condition \eqref{tr1.2aa}, where
\begin{equation}
{\mathcal E}^{(2)}_\nu(\bmod~\omega)=-\frac{1}{T}S(T,\hbar)+
\hbar\sum_{k=1}^n \Omega_k\Big(\nu_k + \frac12\Big).\label{tr2.39}
\end{equation}
The constants ${\mathfrak C}^\nu_T$ are determined by the equation
\begin{equation}
{\mathfrak g}(t,{\mathfrak C}^\nu_T)\Big|_{t=s}={\mathfrak
g}_0(\psi_0)+O(\hbar^{3/2})=\frac{1}{\|\psi\|^2}\langle\psi_0|
\hat{\mathfrak g}|\psi_0\rangle+
O(\hbar^{3/2}),\label{shapovalov:EVOLUT3}
\end{equation}
where
\begin{equation}
\psi_0(\vec x,\hbar)=\Phi^{(2)}_\nu(\vec x,0,\hbar),
\label{phaz_lst4.17}
\end{equation}
and $\Phi^{(2)}_\nu(\vec x,t,\hbar)$ are defined in
\eqref{phaz_lst4.16}.
\end{teo}

By the periodicity condition (\ref{tr2.2a}) for the solutions of
the Hamilton--Ehrenfest system and quasiperiodicity condition
(\ref{tr2.3}) for the solutions of the system in variations it
follows that
\begin{eqnarray}
&\hat a_k(t+T) = e^{i\Omega_kT}\hat a_k(t), \qquad \hat
a(t)=\dfrac1{\sqrt{2\hbar}}\langle a(t),J\Delta\hat z\rangle
\quad k = \overline{1,n},\label{kak6a}\\
&\displaystyle\Big[\hat{\vec
a}^+(t+T)\Big]^\nu=\exp\Big[-i\sum^n_{k=1}T \Omega_k\nu_k\Big]
\Big[\hat{\vec a}^+(t)\Big]^\nu,
\label{tr2.3b} \\
&\displaystyle\det
C(t+T)=\exp\Big[i\sum^n_{k=1}T\Omega_k\nu_k\Big] \det C(t), \qquad
Q(t+T)=Q(t).
\end{eqnarray}

From the relations
\begin{eqnarray}
&&\lefteqn{ |0,t\rangle=|0,t, {\mathfrak
C}_T)\rangle=\displaystyle\frac1{\sqrt{\det
C(t)\sqrt{(n\hbar)^n}}}\times }\cr
&&\quad\displaystyle\times\exp\Big\{\frac i\hbar\Big[S(t,\hbar)+
\langle\vec P(t,\hbar),\Delta\vec x\rangle+\frac
12\langle\Delta\vec x,Q(t)
\Delta\vec x\rangle\Big]\Big\},\label{ros29} \\
&&\displaystyle S(t,\hbar)= \int\limits_0^t dt\Bigl\{\langle\vec
P(t,\hbar)\dot{\vec X}(t,\hbar)\rangle- {\mathcal
H}(z,t)-\tilde\varkappa V_{ww}(z, w,t)\cr &&\quad-
 \frac{\tilde\varkappa}{2}\mbox{Sp}\, \Big[V_{ww}(z,
w,t) }\Delta_{2}(t,{\mathfrak
C}_T)\Big]\Bigr\}\Big|_{z=w=Z(t,{\mathfrak C}_T),  \label{kak15}
\end{eqnarray}
which determine the vacuum trajectory-coherent state, the
definition of semiclassical TCS's \cite{shapovalov:BTS1}
\begin{equation}
\Phi_\nu(\vec x,t,\hbar,{\mathfrak C}_T)=|\nu,t\rangle=|\nu,t,
{\mathfrak C}_T)\rangle=\frac 1{\sqrt{\nu!}} (\hat a^+(t,
{\mathfrak C}_T)))^\nu |0,t, {\mathfrak
C}_T)\rangle=\prod_{k=1}^n\frac 1{\sqrt{\nu_k!}}(\hat a_k^+(t,
{\mathfrak C}_T)))^{\nu_k} |0,t, {\mathfrak
C}_T)\rangle.\label{deist}
\end{equation}
and the quasi-periodicity conditions (\ref{tr2.3b}) it follows
that
\begin{equation}
|\nu,t+T\rangle=e^{- i{\mathcal E}^{(2)}_\nu T/\hbar}
|\nu,t\rangle,\label{e2.36}
\end{equation}
where ${\mathcal E}^{(2)}_\nu$ is defined in \eqref{tr2.39}.
Hence, the functions $\Psi_\nu(\vec x,t,\hbar)$
(\ref{shapovalov:CAUCHY-SOL}) also satisfy the quasi-periodicity
condition (\ref{e2.36}) and constitute the quasi-energy spectral
series $[\Psi_\nu(\vec x,t,\hbar),{\mathcal E}^{(2)}_\nu]$ of
equation \eqref{tr1.1} which correspond to the phase curve
$z=Z_0(t)$.

\section{Semiclassically concentrated solutions of the nonlocal
Gross--Pitaevsky equation (the principal term of the asymptotics)}

The asymptotic solutions of the Cauchy problem
(\ref{phaz_lst1.9})--(\ref{phaz_lst1.12}) (where $\psi_0$ has the
form of (\ref{phaz_lst4.17})) that have been constructed in the
previous section are a special case of the semiclassically
concentrated $(\bmod\hbar^{3/2})$ solutions of equation
(\ref{tr1.1})--(\ref{phaz_lst1.3a}). In case of arbitrary initial
conditions $\psi_0(\vec x,\hbar)$ (\ref{phaz_lst1.12}) belonging
to the class ${\mathcal
P}_\hbar^t\big(Z^{(2)}(t,\hbar),S^{(2)}(t,\hbar)\big)$, $t=s$, the
principal term of the asymptotics $\Psi_0^{(2)}(\vec x,t,\hbar)$
of the nonlocal Gross--Pitaevsky equation
(\ref{tr1.1})--(\ref{phaz_lst1.3a}) is determined by expanding the
solutions of equation (\ref{phaz_lst3.14}) in a series over the
Fock basis $|\nu,t, {\mathfrak C}_T)\rangle$ (\ref{phaz_lst4.16}),
$\nu\in{\mathbb Z}_+^n$:
\[ \Psi_0^{(2)}(\vec x,t,\hbar)=\sum_{|\nu|=0}^\infty C_\nu|\nu,t,
  {\mathfrak C}_T(\psi_0)\rangle, \quad C_\nu=
  \langle {\mathfrak C}_T(\psi_0),0,\nu|\psi_0\rangle. \]
Constants ${\mathfrak C}_T(\psi_0)$ are determined by the equation
\eqref{shapovalov:CONS-2}. For the subsequent calculations, it is
convenient to represent the solution $\psi_0^{(2)}(\vec
x,t,\hbar)$ in terms of the convolution of the initial condition
$\psi_0$ with the Green's function $G_0^{(2)}(\vec x,\vec y,t,s,
{\mathfrak C}_T))$ of equation  (\ref{phaz_lst3.14}). The Green's
function for quadratic quantum systems is well known (see, e.g.,
\cite{Manko, Dodonv76}). For completeness we give its explicit
form in a convenient representation allowing us to clearly
demonstrate the nontrivial dependence of the evolution operator of
the associated linear equation (\ref{phaz_lst3.14}) on the initial
conditions for the input nonlocal Gross-Pitaevsky equation.

By the definition of Green's function $G_0^{(2)}$, for equation
(\ref{phaz_lst3.14}) we have
\begin{equation}
\begin{array}{c}
[-i\hbar\partial_t + \hat{\mathfrak H}_0 (t,{\mathfrak C}_T))]
G_0^{(2)}(\vec
x,\vec y,t,s, {\mathfrak C}_T))=0, \qquad 0\le s\le t;\\[4pt]
\lim\limits_{t\to s}G_0^{(2)}(\vec x,\vec y,t,s, {\mathfrak
C}_T))=\delta(\vec x-\vec y),
\end{array}\label{phaz_lst5.1}
\end{equation}
where the operator $\hat{\mathfrak H}_0$ is defined in
(\ref{phaz_lst3.15}). Denote by $\lambda_k(t, {\mathfrak C}_T)$,
$k=1,2,3,4$, the $n\times n$ matrices that are blocks of the
fundamental matrix of the system in variations
(\ref{phaz_lst4.1}):
\begin{equation}
\Phi(t, {\mathfrak C}_T))=\begin{pmatrix} \lambda_4^\intercal(t,
{\mathfrak C}_T)) & \lambda_2^\intercal(t,{\mathfrak C}_T))\\
\lambda_3^\intercal(t, {\mathfrak C}_T)) & \lambda_1^\intercal(t,
{\mathfrak C}_T))\end{pmatrix}, \qquad \Phi(0, {\mathfrak
C}_T))={\mathbb I}_{2n\times 2n}.\label{phaz_lst5.2}
\end{equation}
Let the following conditions be fulfilled:
\begin{equation}
\det{\mathfrak H}_{pp}(t, {\mathfrak C}_T)) \neq 0, \quad \det
\lambda_3(t-s, {\mathfrak C}_T)\neq 0, \qquad
s,t\in[0,T],\label{phaz_lst5.3}
\end{equation}
Then the Green's function $G_0^{(2)}(x,y,t,s, {\mathfrak C}_T))$
has the form
\begin{gather}
G_0^{(2)}(\vec x,\vec y,t,s, {\mathfrak C}_T))=\frac
  1{\sqrt{\det(-i2\pi\hbar\lambda_3(\Delta t, {\mathfrak C}_T))}} \exp\Big\{\frac i
  \hbar\Big[S^{(2)}(t,\hbar)-S^{(2)}(s,\hbar)+\nonumber\\
+\langle \vec P^{(2)}(t,\hbar),\Delta \vec x\rangle-\langle \vec
  p_0,(\vec y-\vec x_0)\rangle-\frac12\langle(\vec y- \vec x_0),
  \lambda_1(\Delta t, {\mathfrak C}_T)\lambda^{-1}_3(\Delta t, {\mathfrak C}_T)(\vec y-\vec
  x_0)\rangle+\nonumber\\
+\langle\Delta \vec x,\lambda_3^{-1}(\Delta t, {\mathfrak C}_T)(
\vec y- \vec x_0)
  \rangle-\frac12\langle\Delta\vec x,\lambda_3^{-1}(\Delta t, {\mathfrak C}_T)\lambda_4
  (\Delta t, {\mathfrak C}_T)\Delta\vec x\rangle\Big] \Big\}. \label{phaz_lst5.4}
\end{gather}
Here, $\Delta t=t-s$, $Z^{(2)}(t,\hbar)=(\vec P^{(2)}(t,\hbar),
\vec X^{(2)}(t,\hbar))$ is the projection of the solution of the
second order Hamilton--Ehrenfest system onto ${\mathbb
R}^{2n}_{p,x}$.

\begin{rem}
If condition \eqref{phaz_lst5.3} is not fulfilled, the solution of
the problem \eqref{phaz_lst5.1} has a somewhat different form than
\eqref{phaz_lst5.4} (see, e.g., \cite{Dodonv76}).
\end{rem}

From this reasoning and Statement \ref{d1}\ we arrive at the
following theorem.

\begin{teo} \label{theor-5.1} Let the symbols of the operators
$\widehat{\mathcal H}(t)$ and $\widehat V(t,\Psi)$ in
\eqref{tr1.1}--\eqref{phaz_lst1.3a} satisfy the conditions of
Assumption ~\ref{assum1} and let the conditions of Theorem
\ref{teor-4.1}\ and conditions \eqref{phaz_lst5.3} be fulfilled.
Then the function
\begin{equation}
\psi_0^{(2)}(\vec x,t,\hbar)=\widehat U_0^{(2)}(t,0)\psi_0, \quad
t\in[0,T], \label{phaz_lst5.5}
\end{equation}
where $\widehat U_0^{(2)}(t,0)$ is the evolution operator of the
zero order associated Schr\"odinger equation \eqref{phaz_lst3.14}
with a kernel $G_0^{(2)}(\vec x,\vec y,t,0, {\mathfrak
C}(\psi_0)))$ \eqref{phaz_lst5.4}, is an asymptotic $($to within
$O(\hbar^{3/2})$, $\hbar\to0)$ solution of the nonlocal
Gross--Pitaevsky equation \eqref{tr1.1}--\eqref{phaz_lst1.3a} with
the initial condition
\[ \Psi|_{t=0}=\psi_0(\vec x).\]
\end{teo}
\begin{sled} Operator $\widehat U_0^{(2)}(T)=\widehat U_0^{(2)}(T,0)$
 is the monodromy
operator the Gross--Pitaevsky equation
\eqref{tr1.1}--\eqref{phaz_lst1.3a}. Functions $\Psi_\nu(\vec
x,t,\hbar)$ \eqref{shapovalov:CAUCHY-SOL} are eigenfunctions of
operator $\widehat U_0^{(2)}(T)$:
\[\widehat U_0^{(2)}(T)\Psi_\nu(\vec x,t,\hbar)
=e^{- i{\mathcal E}^{(2)}_\nu T/\hbar}\Psi_\nu(\vec x,t,\hbar).\]
\end{sled}

\section{Geometric phases of trajectory-coherent states }

We now turn to the calculation of the Aharonov--Anandan phase
corresponding to the quasi-energy states (\ref{tr2.39}). To do
this, we use formula (\ref{tr1.11}) by which, neglecting the terms
of order $O(\hbar)$, we obtain the following expression of the
desired phase if the phase curve is periodic:
\begin{equation} 
\gamma_{{\mathcal E}_\nu}=-\frac{{\mathcal E}^{(2)}_\nu T}\hbar+
\frac 1\hbar\int\limits_0^T dt\bigg\{{\mathcal H}(t)
+\tilde\varkappa V(t)+\frac 12 \mbox{Sp}\Big( {\mathfrak
H}_{zz}(t)\Delta_2\Big) 
+\frac{\tilde\varkappa}2 \mbox{Sp}\,\bigr[
V_{\omega\omega}(Z(t,\hbar),\omega,t)\Delta_2\Bigl|_{\omega=Z(t,\hbar)}\bigr]
\bigg\} .\label{lisok1}
\end{equation}

Using (\ref{kvs12a}), we can readily show that
\begin{gather}
\mbox{Sp}\Big([{\mathcal H}_{zz}(t) + \varkappa
  V_{zz}(t)]\Delta_2(t)\Big)=\frac{\hbar}{2}~{\rm Sp~Re}\, \Big[\dot
  C(t)D^{-1}_\nu B^+(t)-\dot B(t)D^{-1}_\nu C^+(t)\Big]=\nonumber\\
=-\hbar~\mbox{Re}\sum_{k=1}^n\Big(\nu_k+\frac12\Big)\{\dot
a_k(t),a^*_k(t)\}, \label{phazcyc3.4}
\end{gather}
where, in our case, $D^{-1}_\nu=\mbox{diag}
(2\nu_1+1,\dots,2\nu_n+1)$. Note that the quantity under the
summation sign on the right of (\ref{phazcyc3.4}), by virtue of
(\ref{kvs10}), is real. Hence, the sign Re can be omitted.
Introduce the notation $a_0(t)=\dot Z_0(t)$, where
$Z(t,\hbar)=Z_0(t)+\hbar Z_1(t) + O(\hbar^{3/2})$ are defined in
(\ref{kvs7}). Then, substituting the explicit expression for
quasi-energies $\mathcal {E}^{(2)}_\nu$ (\ref{tr2.39}) in
(\ref{lisok1}) and using relation (\ref{phazcyc3.4}), we find
\begin{equation}
\gamma_{{\mathcal E}_\nu}=\frac1\hbar \int\limits_0^T dt
\langle\vec P(t),\dot{\vec
X}(t)\rangle-\int\limits_0^T dt\{\dot Z_0(t),Z_1(t)\} 
-\sum_{k=1}^n \Big(\nu_k + \frac12\Big)\bigg[ T\Omega_k +
\frac{1}{2} \int\limits_0^T dt\{\dot
a_k(t),a^*_k(t)\}\bigg].\label{phazcyc3.5}
\end{equation}
If now we introduce, instead of the Floquet solutions, the
$T$-periodic vector functions
\begin{equation}
\tilde a_k(t) =e^{-i\Omega_kt}a_k(t), \qquad \tilde a_k(t+T) =
\tilde a_k(t), \label{phazcyc3.6}
\end{equation}
then formula (\ref{phazcyc3.5}) takes the form
\begin{equation}
\gamma_{{\mathcal E}_\nu}=\frac1\hbar\int\limits_0^T dt
\langle\vec P(t),\dot{\vec X}(t)\rangle-\int\limits_0^T
dt\{a_0(t),Z_1(t)\}
-\frac12\sum_{k=1}^n \Big(\!\nu_k+\frac12\!\Big)\int\limits_0^T dt
\{\dot{\tilde a}_k(t),{\tilde a}^*_k(t)\}.\label{phazcyc3.7}
\end{equation}

Using (\ref{kvs11a}), we obtain the following expression for the
Aharonov--Anandan phase (\ref{phazcyc3.7}):
\begin{align}
\gamma_{{\mathcal E}_\nu}&=\frac1\hbar \int\limits_0^T dt
\langle\vec P(t),\dot{\vec X}(t)\rangle -
\frac{1}{2\hbar}\sum_{k=1}^n \Big(\nu_k +
\frac12\Big)\int\limits_0^T dt\{ \dot{\tilde a}_k, {\tilde
a}^*_k\} - \nonumber\\
& -\mbox{Re}\int\limits_0^T dt\frac{i}{2}
\sum_{l=1}^n\{a_0(t),{\mathfrak a}_l(t)\} \biggl[
\!\sum_{k=1}^n\Big(\nu_k+\frac12\Big) \int\limits_0^t
d\tau\{{\mathfrak a}^*_l(\tau),\partial_z\}\Big\{a^*_k(\tau),J
\Big[{\mathcal H}_{zz}(z,\tau)+\nonumber\\
& +\tilde\varkappa V_{zz}(z,w,\tau)+ \tilde\varkappa
V_{ww}(z,w,\tau)\Big]a_k(\tau)\Big\}\!\bigg]\Big|_{z=w=Z_0(\tau)}+
b_k(T)e^{i{\widetilde \Omega}_kT}\biggr].\label{phazcyc3.18}
\end{align}

Thus, the the Aharonov--Anandan phase $\gamma_{{\mathcal E}_\nu}$
corresponding to the quasi-energy TCS's $\Psi_{{\mathcal E}_\nu}$
(\ref{shapovalov:CAUCHY-SOL}) is completely determined by two
geometric objects: the closed phase trajectory $Z_0(t)$ of the
Hamilton--Ehrenfest system, stable in the linear approximation,
and the complex germ $r^n(Z_0(t))$ composed of $n$ linearly
independent Floquet solutions of the system in variations
(\ref{kvs9}).



Now consider Eqs. (\ref{tr1.1})--(\ref{phaz_lst1.3a}) in a
3-dimensional space with operators $\widehat{\mathcal H}(t)$,
$\widehat V(t,\Psi)$ of the form
\begin{eqnarray}
&{\mathcal H}(\hat z,t)=\dfrac{1}{2m} \hat{\vec p}^2-e \langle
\vec E(t),\vec x\rangle+\dfrac{k}{2}\vec x^2, \label{shapovalov:EXAMP1-1}\\
& V(\hat z,\hat w,t) =V(\vec x-\vec y)= V_0\exp\Big[ -\dfrac{(\vec
x-\vec y)^2}{2\gamma^2}\Big]. \label{shapovalov:EXAMP1-2}
\end{eqnarray}
The external field  in the linear operator
(\ref{shapovalov:EXAMP1-1}) is  electric field $\vec
E(t)=(E\cos\omega t,E\sin\omega t,0)$ periodic in time with
frequency $\omega$, and the field of an isotropic oscillator with
potential $k\vec x^2/2$, $k>0$.

The periodic solution (\ref{kvs8}) corresponding to
(\ref{shapovalov:EXAMP1-1}),(\ref{shapovalov:EXAMP1-2}) accurate
to $O(\hbar^{3/2})$
\begin{eqnarray}
&&Z_0(t,\hbar)=(\vec P_0(t,\hbar), \vec X_0(t,\hbar)), \label{shapovalov:2.3}\\
&&\vec P_0(t,\hbar)=\Big(-m \omega\xi\sin\omega t, m \omega\xi\cos\omega t, 0 \Big), \nonumber\\
&&\vec X_0(t,\hbar)=\big(\xi\cos\omega t, \xi\sin\omega t, 0\big).
\quad 
\nonumber 
\end{eqnarray}
Here $\xi=eE/[m(\omega_0^2-\omega^2)]$.

The Floquet solutions of system (\ref{kvs9}) corresponding to
(\ref{shapovalov:EXAMP1-1}), (\ref{shapovalov:EXAMP1-2}) accurate
to $O(\hbar^{3/2})$:
\begin{eqnarray}
&&a_1(t)=e^{i\omega_{\rm s}t}\Big(g_{\rm s},0,0,-
  \dfrac{i}{g_{\rm s}},0,0\Big)^\intercal, \nonumber\\
&&a_2(t)=e^{i\omega_{\rm s}t}\Big(0,g_{\rm s},0,0,
  -\dfrac{i}{g_{\rm s}},0)^\intercal, \label{shapovalov:2.7}\\
&&a_3(t)=e^{i\omega_{s}t}\Big(0,0,g_{\rm s},0,0,
  -\dfrac{i}{g_{\rm s}})^\intercal.\nonumber
\end{eqnarray}
Here
 \[ 
\omega_{\rm s}=\sqrt{\omega_0^2-\eta\omega_{\rm nl}^2};\quad
g_{\rm s}=\sqrt{m\omega_{\rm s}};\quad \omega_{\rm
nl}=\sqrt{\frac{|\tilde\varkappa V_0|}{m\gamma^2}};\quad
\omega_0=\sqrt{\frac{k}{m}};\quad
\tilde\varkappa=\varkappa\|\Psi\|^2;\quad
\eta=\mbox{sign}\,(\tilde\varkappa V_0). 
\]
 The spectrum of quasi-energies ${\mathcal E}^{(2)}_\nu$ (\ref{tr2.39})   is given by the
relation
\begin{align} \label{shapovalov:3.6}
{\mathcal E}^{(2)}_\nu&=-\frac{eE}{2}\xi+\tilde\varkappa V_0
+\hbar\Big[\Big(\omega_{\rm s}-\frac{\eta\omega_{\rm
nl}^2}{2\omega_{\rm s}}\Big)\Big(\nu_1+\frac 12\Big)+\nonumber\\
&+\Big(\omega_{\rm s}-\frac{\eta\omega_{\rm nl}^2}{2\omega_{\rm
s}}\Big)\Big(\nu_2+\frac 12\Big)+\Big(\omega_{\rm s}-
\frac{\eta\omega_{\rm nl}^2}{2\omega_{\rm s}}
\Big)\Big(\nu_3+\frac 12\Big)\Big]+O(\hbar^{3/2}).
\end{align}

The Aharonov--Anandan geometric phase (\ref{phazcyc3.18}) is:
\begin{equation}\label{shapovalov:3.9}
\gamma_{{\mathcal E}_\nu}=\frac{T\omega^2}{\hbar}m\xi^2+
O(\hbar^{1/2}).
\end{equation}

We have found the explicit expressions for the Aharonov--Anandan
phase in semiclassical approximation accurate to $O(\hbar^{1/2})$
for the quasi-energy states which are governed by the GPE and
belong to the class of trajectory concentrated functions. It is of
interest to consider the expressions (\ref{phazcyc3.18}) in the
adiabatic limit $T\to\infty$ and to observe in what sense the
Aharonov--Anandan phase transforms to the Berry phase in the
background of formalism of the GPE. The fact is that the splitting
of the total phase, gained by a wave function, into dynamic and
geometric parts is carried out by analogy with the case of linear
quantum mechanics. Then it is not obvious that the expressions
obtained for the nonlinear GPE are determined only by geometry of
the system and do not contain a dynamic contribution due to the
nonlinear term in the equation.

In the limit $T\to\infty$ we can neglect the dependence of
operators $\widehat{\mathcal H}(t)$ and $\widehat V(t,\Psi)$ of
the form (\ref{tr1.1}) on time $t$. Therefore, the quasi-energy
spectral series (\ref{shapovalov:CAUCHY-SOL}) grade into discrete
spectral series of the nonlinear problem
\[ [\widehat{\mathcal H}+\varkappa\hat V (\Psi)]\Psi=E\Psi \]
in the limit $T\to\infty$. These spectral series \cite{BLT07} are
localized in a neighborhood of stationary solutions of the
Hamilton--Ehrenfest system (\ref{phaz_lst2.7}), (\ref{kvs8}). The
geometrical phases (\ref{shapovalov:3.9}) do not contain quantum
corrections accurate to $O(\hbar^{1/2})$ in the considered example
and do not depend on nonlinear potential. Evidently, it is related
with the fact that the nonlinearity and the quantum corrections
with the given accuracy in $\hbar$ do not change solutions
(\ref{shapovalov:2.3}) of the Hamilton system (\ref{kvs8}) and, as
a result, the geometry of the system.

\subsection*{Acknowledgements}
The work was supported in part by  President of the Russian
Federation, Grant  No NSh-5103.2006.2, MK-4629.2006.2 and  DVG,
A. Lisok was supported in part by the scholarship of the nonprofit
Dynasty Foundation.


\begin{thebibliography}{99}

\bibitem{shapovalov:BERRY} Berry~M.V. 1984 Quantum Phase Factors
Accompanying Algebraic Changes, {\em Proc. Roy. Soc. London.} {\bf
A392:~1802} 45-58.
\bibitem{shapovalov:VINI} Vinitskii~S.I., Derbov~V.L., Dubovik~V.M.,
Markovskii~B.L, and Stepanovskii~Yu.P. 1990 Topological Phases in
Quantum Mechanics and Polarized Optics, {\it Uspekhi Fiz. Nauk.
Rus. Acad. Sci.} {\bf160} 1-49 (in Russian).
\bibitem{shapovalov:KLYSHKO} Klyshko~D.N. 1993 Geometric Phase
in Oscillation Processes, {\it Uspekhi Fiz. Nauk. Rus. Acad. Sci.}
{\bf163} 1-18 (in Russian).
\bibitem{Moore} Moore D.J. 1991  The calculation of nonadiabatic Berry
phase, {\em Phys. Rep.}  {\bf 210:1} 1-43.

\bibitem{shapovalov:BTS1} Belov~V.V., Trifonov~A.Yu., and
Shapovalov~A.V. 2002 The Trajectory-Coherent Approximation and the
System of Moments for the Hartree Type Equation, {\em Int. J.
Math. and Math. Sci.} {\bf32:6} 325-370.
\bibitem{shapovalov:BTS2} Belov~V.V., Trifonov~A.Yu., and Shapovalov~A.V. 2002
Semiclassical Trajectory-Coherent Approximations to Hartree Type
Equations, {\em Teor.  Mat. Fiz.} {\bf 130:3} 460-92;\\
Belov~V.V., Trifonov~A.Yu., and Shapovalov~A.V. 2002 {\em Theor.
Math. Phys.} {\bf 130:3} 391-418 (Engl. Transl.).
\bibitem{lst3} Lisok~A.L., Trifonov~A.Yu., and Shapovalov A.V. 2004 The
evolution operator of the Hartree-type equation with a quadratic
potential, {\em J. Phys. A.}  {\bf37} 535-456.

\bibitem{lst7} Lisok A.L., Trifonov A.Yu. and Shapovalov~A.V.
2005  Symmetry operators of a Hartree-type equation with a
quadratic potential. {\em Sibirsk. Mat. Zh.} {\bf 46:1} 149-65
(in Russian).

\bibitem{lst_sigma} Shapovalov~A.V., Trifonov~A.Yu., and   Lisok~A.L. 2005 Exact
Solutions and Symmetry Operators for the Nonlocal Gross-Pitaevskii
Equation with Quadratic Potential {\it Sym., Integ. and Geom.:
Meth. and Appl.} {\bf 1:007} 1-14. 

\bibitem{shapovalov:ZELDOVICH} Zeldovich~Ya.B. 1966
Quasienergy of Quantum System under Periodical Influence, {\it Zh.
Eksper. Teoret. Fiz.} {\bf51:5(11)} 1492-1495 (in Russian).

\bibitem{shapovalov:RITUS} Ritus~V.I. 1966 Atomic Level Shift and Splitting
by the Electromagnetic Field, {\it Zh. Eksper. Teoret. Fiz.}
{\bf51:5(11)} 1544-1549 (in Russian). 

\bibitem{shapovalov:ANANDAN} Anandan J. and Aharonov Y. 1988 Geometric
Quantum Phase and Angles, {\em Phys. Rev. D}. {\bf38:6} 1863-1870.

\bibitem{Manko} Malkin M.A. and Man'ko V.I. 1979 {\em Dynamic Symmetries and Coherent
States of Quantum Systems}. (Moscow: Nauka) (in Russian).

\bibitem{Chern}  Perelomov A. M. 1986 {\em Generalized Coherent States and Their Applications}. (Berlin:
Springer).


\bibitem{shapovalov:BAGRE} Bagrov~V.G., Belov~V.V., and
Trifonov~A.Yu. 1996 Semiclassical Trajectory-coherent
Approximation in Quantum Mechanics: I. High Order Corrections to
Multidimensional Time-dependent Equations of Schr\"odinger Type,
{\em Ann. Phys.} (NY), {\bf246:2} 231-280. 

\bibitem{shapovalov:KARASEVMASLOV} Karasev M.V. and Maslov V.P. 1993
{\em Nonlinear Poisson Brackets: Geometry and Quantization}, 
Ser. Traslations of Mathematical Monographs, Vol. 119, Amer. Math.
Soc., Providence, RI. 

\bibitem{vs2} Trifonov A.Yu. and Yevseyevich A.A. 1995 The
Aharonov--Anandan Phase for Quasi-energy Trajectory-Coherent
States, {\em J. Phys. A: Math. Gen.} {\bf28} 5653-5672.

\bibitem{88a} Bagrov V.G., Belov V.V., Kondratyeva M.F., Rogova~A.M.,
and Trifonov A.Yu. 1993 A New Formulation of Quasi-classical
Approximation in Quantum Mechanics, {\em J. Moscow Phys. Soc.}
{\bf3} 1-12. 

\bibitem{Robertson} Robertson H.P. 1934 An Indeterminacy Relation for
Several Observables and its Classical Interpretation, {\em  Phys.
Rev.}  {\bf46: 9} 794-801. 

\bibitem{DoMa1} Dodonov V.V. and Man'ko V.I. 1985 Universal Invariants of Quantum
Systems and Generalized Uncertainty Relation, {\em Group
Theoretical Methods in Physics} {bf 1} (London, Paris, New York:
Harwood Acad. Publ.) 

\bibitem{shapovalov:MAS}  Maslov V.P. 1977 {\em The Complex WKB Method in Nonlinear
Equations} (Moscow: Nauka), in Russian; Maslov V.P. 1994 {\em The
Complex WKB Method for Nonlinear Equations}. I. {\em Linear
Theory} 1994 (Basel,Boston, Berlin: Birkhauser Verlag) (Engl.
Transl.)

\bibitem{BeD2} Belov V. V. and Dobrokhotov S. Yu. 1992
Semiclassical Maslov Asymptotics with Complex Phases. I. General
Approach {\em Teoret. Mat. Fiz.} {\bf 92:2} 215--54 (in Russian).

\bibitem{bbt82} Bagrov V. G., Belov V. V., and Ternov I. M. 1982 Quasiclassical
Trajectory-Coherent States of a Nonrelativistic Particle in an
Arbitrary Electromagnetic Field, {\em Teoret. Mat. Fiz.}
{\bf 50:3} 390--96 (in Russian). 

\bibitem{bbt83} Bagrov V.G., Belov V.V., and Ternov I.M. 1983 Quasiclassical
trajectory-coherent states of a particle in arbitrary
electromagnetic field, {\em J. Math. Phys.} {\bf24} 2855-2859.


\bibitem{Arnold2} Arnold V.I. 1989 {\em Mathematical Methods in Classical Mechanics}.
(Moscow: Nauka) (in Russian). 

\bibitem{Simon} Reed M. and Simon B. 1972 {\it Methods of Modern Mathematical Physics},
Vol. 1. (New York, London: Acad. Press). 

\bibitem{Dodonv76} Dodonov V.V., Malkin I.A., and Man'ko V.I. 1975 Integrals of motion,
Green functions and coherent states of dynamic systems, {\em
Intern. J. Theor. Phys.} {\bf 14: 1} 37-54. 

\bibitem{BLT07} Belov V. V., Litvinets F. N. and Trifonov A. Yu. 2007 The semiclassical
spectral series for a Hartree-type equation corresponding to a
rest point of the Hamilton–Ehrenfest system, {\em Teor.  Mat. Fiz.} {\bf  150:1} 26-40;\\
Belov V. V., Litvinets F. N. and Trifonov A. Yu. 2007 {\em Theor.
Math. Phys.} {\bf 150:1} 21-33 (Engl. Transl.).

\end{thebibliography}
\end{document}